\documentclass[useAMS,usenatbib]{mn2e}

\usepackage{amsmath}
\usepackage{amssymb}
\usepackage{epsfig}
\usepackage{ifthen}
\usepackage{latexsym}
\usepackage{rotating}
\usepackage{subfigure}
\usepackage{times,epsf}
\usepackage{txfonts}
\usepackage{varioref}
\usepackage{verbatim}
\usepackage{url}
\usepackage{color}
\usepackage{array}
\usepackage[dvipsnames]{xcolor}
\usepackage[T1]{fontenc}
\usepackage{multirow}
\usepackage{graphicx}

\newcolumntype{M}{>{$\vcenter\bgroup\hbox\bgroup}c<{\egroup\egroup$}}

\newcommand{\be}{\begin{equation}}
\newcommand{\ee}{\end{equation}}
\newcommand{\bea}{\begin{eqnarray}}
\newcommand{\eea}{\end{eqnarray}}

\newcommand\apj{Astrophysical Journal}
\newcommand\apjl{Astrophysical Journal Letters}

\newcommand\aap{Astronomy \& Astrophysics}

\newcommand\nat{Nature}

\newcommand\mnras{Monthly Notices of the Royal Astronomical Society}

\newcommand\procspie{Proceedings of the International Society for Optical Engineering}

\newcommand{\DA}[1]{#1}
\newcommand{\sgra}{Sgr~A$^*$ }
\newcommand{\sgrano}{Sgr~A$^*$}
\newcommand{\Rg}{\,R_{\rm G}}

\title[Effects of \sgra on G2]{Simulating the effect of the \sgra accretion flow on the appearance of G2 after pericenter}

\author[D. Abarca, A. S\k{a}dowski, L. Sironi] {David Abarca$^{1}$\footnotemark[1], Aleksander
  S\k{a}dowski$^1$\footnotemark[1], Lorenzo Sironi$^{1,2}$\thanks{E-mail: dabarca@cfa.harvard.edu (DA); asadowski@cfa.harvard.edu (AS); lsironi@cfa.harvard.edu (LS)} \\ $^1$ Harvard-Smithsonian Center for
  Astrophysics, 60 Garden St., Cambridge, MA 02134, USA\\
  $^2$ NASA Einstein Fellow}

\begin{document}

\maketitle

\label{firstpage}

\begin{abstract}
  We study the dynamical interaction of the G2 cloud with the  accretion flow around \sgra
 by means of three-dimensional, hydrodynamic
  simulations. We show the effects of the rotating accretion flow on
  the evolution of G2 by projecting the cloud density onto the plane of the sky, and
  extracting position-velocity diagrams. We study a number of possible
  orientations of the cloud orbit with respect to the disk. We find
  that once the center of mass of the cloud has crossed the pericenter, the differences
  between models becomes significant. Models with the cloud counter-rotating with respect to
  the disk are expected to reach higher blue-shifted line of sight
  velocities. The spatial extent of the emission depends strongly on
  the cloud-to-disk inclination angle. Future imaging
  and spectroscopy of G2 emission will shed light both on the structure
  of the \sgra disk and on the properties of the cloud.
\end{abstract}


\section{Introduction}
\label{s.introduction}


\cite{gillessen+12a} discovered a gaseous object (the G2 cloud)
approaching the supermassive black hole (SMBH) in the center of the Galaxy on
a very eccentric orbit. It is expected to approach as close as $R=4800\Rg$
($R_{\rm G}=GM_{\rm BH}/c^2=6.3\cdot 10^{11}\,\rm cm$ is the black hole gravitational radius, where $M_{\rm BH}$ is the black hole mass) from \sgra in early 2014 \citep{gillessen+13}, and its
mass has been estimated to be roughly $3M_\oplus$
\citep{gillessen+12a}. The nature of the cloud is still
debated. A pressure supported gas cloud or a momentum supported wind  from
a star (where
the external 
pressure is balanced by its ram pressure) may explain equally well the observed size, shape, and range
of line of sight velocities \citep{schartmann+12}.  Although, the fact
that the cloud is followed by an additional structure (`tail') of comparable mass 
and very similar orbit, \cite{gillessen+12a, burkert+12} support the scenario of a gas cloud which
was formed not so long ago in the Galactic center.

On its way, the cloud evolves under the tidal gravitational field of the
SMBH. The resulting stretching has been clearly observed in combined
astrometric and spectroscopic data. The pressure supported gas cloud
model predicts that the object will be tidally disrupted after crossing the pericenter,
while the stellar wind model predicts that the object survives the
passage. However, the tidal forces are not the only ones acting on
the cloud. \sgra is known to accrete gas with a relatively low accretion
rate of $10^{-9}-10^{-7} M_\odot/{\rm yr}$ (see next
Section). However, even such a small rate
implies an accretion flow with densities and velocities in its inner
region comparable to those of G2. Therefore, the cloud itself is likely to be affected by
its dynamical interaction with the accretion disk of \sgrano. The outcome
will depend on the properties of the cloud and the accretion flow and
orientation of the cloud orbit with respect to the accretion disk
axis. The point of this paper is to study how this
interaction affects the observables around and after pericenter
assuming that G2 is a pressure supported gas cloud.

The G2 cloud has so far been observed by two groups with adaptive
optics astrometry and spectroscopy based on the Keck and VLT
telescopes \citep{gillessen+12a,gillessen+12b,gillessen+13,keckg2}. They were able to extract not only images of the cloud approaching the
Galactic center but also the position along the orbit (arclength along the center of mass trajectory to the
pericenter) vs line of sight
velocity diagrams, which allowed for very precise determination of
the orbital parameters of the cloud center of mass. As the cloud becomes tidally
elongated, the observed emission no longer follows a single
Keplerian orbit \citep{gillessen+13}. The spatial distribution and
line of sight properties of the emission are likely to be affected by
dynamical interactions with the accretion flow and to trace the
properties of \sgrano's rotating atmosphere. The interaction will
likely also affect the plane of sky view of the cloud. Therefore, it
is not unreasonable to think that the observations of the cloud near
and after the epoch of pericenter will shed some light on the so far
poorly constrained properties of the \sgra accretion disk and/or the G2 cloud.

These hopes are especially justified in the view of the soon to come
leading infrared imaging project, GRAVITY.  It is a second generation
VLTI, four beam combiner with 4 miliarcsecond (mas) imaging resolution
for a $m_K\sim15$ source \citep{kendrew+gravity,gillessen+gravity}.
First astronomical light for this instrument is planned for 2014,
possibly early enough to witness the tail end of G2 pericenter
passage and to contribute to
the imaging of G2 after it has crossed the pericenter.

\subsection{Previous work}
\label{s.previous}

Much work has already been done on simulating G2's interaction with \sgra. 
\cite{schartmann+12} used two-dimensional, high resolution, 
hydrodynamic simulations to evolve two 
different origin scenarios: a compact cloud scenario with a uniform density structure formed shortly before
first observations, and a 
spherical shell with an apocenter among the disks of young stars in the Galactic Center.
They modeled the atmosphere of \sgra by neglecting  rotation and 
choosing the temperature to maintain hydrostatic 
equilibrium.

\cite{anninos+12} performed three-dimensional, moving-mesh,
hydrodynamic simulations.
They assumed a spherical
cloud in pressure equilibrium as in \cite{schartmann+12}. They used
two different models for the SMBH atmosphere, one following \cite{schartmann+12}
which is convectively unstable, and a more physically
motivated one from \cite{yuan+03} with a density and temperature
profile based on observations, again without rotation.
They adopted an initial temperature of $10^4$ K 
and claimed that the radiative cooling timescale is short enough to maintain this temperature.
They support the conclusions of \cite{schartmann+12} and
\cite{burkert+12} that the 
cloud must have been formed shortly before its first observation. They did not 
expect the cloud to survive its pericenter passage.

\cite{saitoh+12} attempted to address the radiative cooling problem using
N-body/smoothed particle hydrodynamics. Using the same background setup as
\cite{schartmann+12} but scaling the disk density as a parameter, they 
predicted a maximum bolometric luminosity of $\sim 100 L_\odot$ which should be visible
in the near IR. They did not
account for rotation.

\cite{sadowski+G22} studied the formation of the bow-shock in front of
the cloud when penetrating the \sgra accretion disk. They used a
turbulent and magnetized
background model of the disk based on a general relativistic
magneto-hydrodynamic (GRMHD) simulation. The cloud had to be scaled
down to fit within the computational domain. Its density, velocity and mass
 were also scaled to produce a consistently scaled model of the
 entire system. The interaction was
resolved under full GRMHD, yet with very limited resolution in azimuth.

The simulations we report in this work do not neglect the rotation of
the \sgra atmosphere and, for the first time, allow for the study of the impact
of the orientation of the cloud orbit with respect to the axis of
rotation of the atmosphere. As with most of the previous authors, we
neglect the magnetic field. On one hand, resolving the magneto-rotational
instability (MRI) without scaling down the cloud is impossible with
current computational resources, on the other magnetic fields are not likely to substantially
change the nature of the interaction, since they are subdominant with
respect to 
gas pressure.

The main alternative to the cloud scenario, the compact source scenario,
is addressed by \cite{ballone+13} who simulated various setups of G2 as 
an outflow from a central source using two-dimensional simulations in cylindrical coordinates. 
Assuming a zero angular momentum orbit, they tracked the cloud from apocenter until
very near \sgra. They found that their simulations were consistent with
the observed extent of G2 on the position-velocity diagrams.

\DA{\cite{guillochon+14} studied the origin of G2 as the result of a tidal disruption event of
a star orbiting the Galactic center. Using 3D, hydrodynamic code FLASH, \cite{guillochon+14} 
found that the returning gas tidally stripped from a star may 
fragment and explain the clumpy nature of G2 and its observed tail.}

\section{\sgra disk}
\label{s.disk}

The SMBH at the center of our Galaxy is known to accrete gas at fairly
low level of $\dot M_{\rm BH}=10^{-9} - 10^{-7}\,M_{\odot}/\rm yr$
\citep{yuan+03,marrone+07,moscibrodzka+09,dexter+10,roman+12,dibi+12}.
 The accretion flow is radiatively inefficient and its
properties are not yet well determined. The only well-established
parameter
besides the accretion rate at the BH is the density at the Bondi
radius $R_{\rm B}$, $n_{\rm B}=120\,\rm cm^{-3}$, where
$R_{\rm B}=2\times 10^5\,R_{\rm G}$ \citep{baganoff+03}. Recent Chandra
observations of the region inside the Bondi radius suggest the density
slope, $n \propto r^{-0.5}$ \citep{chandra-3ms}, but are far from being
conclusive. To fill the gap between the BH and $R_{\rm B}$ one has to
rely on numerical models of radiatively inefficient accretion flows
(RIAFs).

Simulating accretion across such a wide range of scales (from the BH horizon
to the Bondi radius) is a daunting challenge. Single simulations which
resolve the BH horizon in general relativity are limited by their
computational cost to two decades in radius \citep[e.g.,][]{narayan+12b}. Larger scales may
be resolved but only by means of coupling separate simulations
together \citep{yuan+12a}. 

There is a consensus, based on a limited number of long duration simulations
performed so far, that RIAFs are characterized by significant mass
loss leading to a shallower slope of the density distribution than the slope, $n \propto r^{-3/2}$,
predicted by the standard ADAF model \citep{narayanyi94}. The slope favored by
simulations is close to $n \propto r^{-1}$ and for \sgra it
qualitatively agrees with the measured value of the accretion rate at
the horizon.

In this work we use the \sgra accretion flow model put forward by
\cite{sadowski+G21}. It is based on a long-duration GRMHD simulation
of a RIAF \citep{narayan+12b} which has converged up to $R\lesssim 200R_{\rm
  G}$. The power-law profiles of density (normalized to fit the density at the Bondi
radius), temperature, and azimuthal velocity were extrapolated along
the slopes implied by the data inside the resolved domain. Such a solution provides
axisymmetric accretion flow structure as a function of radius and polar coordinate 
in the whole region between the BH and the
Bondi radius, which reasonably reproduces the \sgra accretion
flow under the assumption that its structure is self-similar there.

While we consider several scenarios for the orientation of the G2 orbit relative to the accretion flow around
\sgra, recent millimeter VLBI observations 
of the Event Horizon Telescope collaboration \citep{broderick+11}
suggest that G2 is roughly co-rotating
with respect to the spin direction of \sgra. There are large uncertainties
on these estimates however. We will consider the co-rotating scenario
as our fiducial model.

\section{Cloud model}
\label{s.cloud}

The G2 cloud has been observed in infrared since 2004. The
combined spectroscopic and imaging data have allowed for a precise
determination of the orbital parameters of the cloud center of mass
\citep{gillessen+12a,gillessen+12b,gillessen+13,keckg2}. 
The position and line of sight velocity measurements have been
successfully reproduced by a model consisting of test particles
initiated at year $t=2000.0$ as a Gaussian spherical cloud with full width
at half maximum (FWHM) of 42 milliarcseconds (mas) with velocity dispersion
$\sigma=120\,\rm km/s$ \citep{gillessen+13}. Although this approach qualitatively
reproduces the observed position and velocity distribution, it has a few
flaws that we address below.

\begin{figure}
  \centering
\subfigure{\includegraphics[width=.9\columnwidth,]{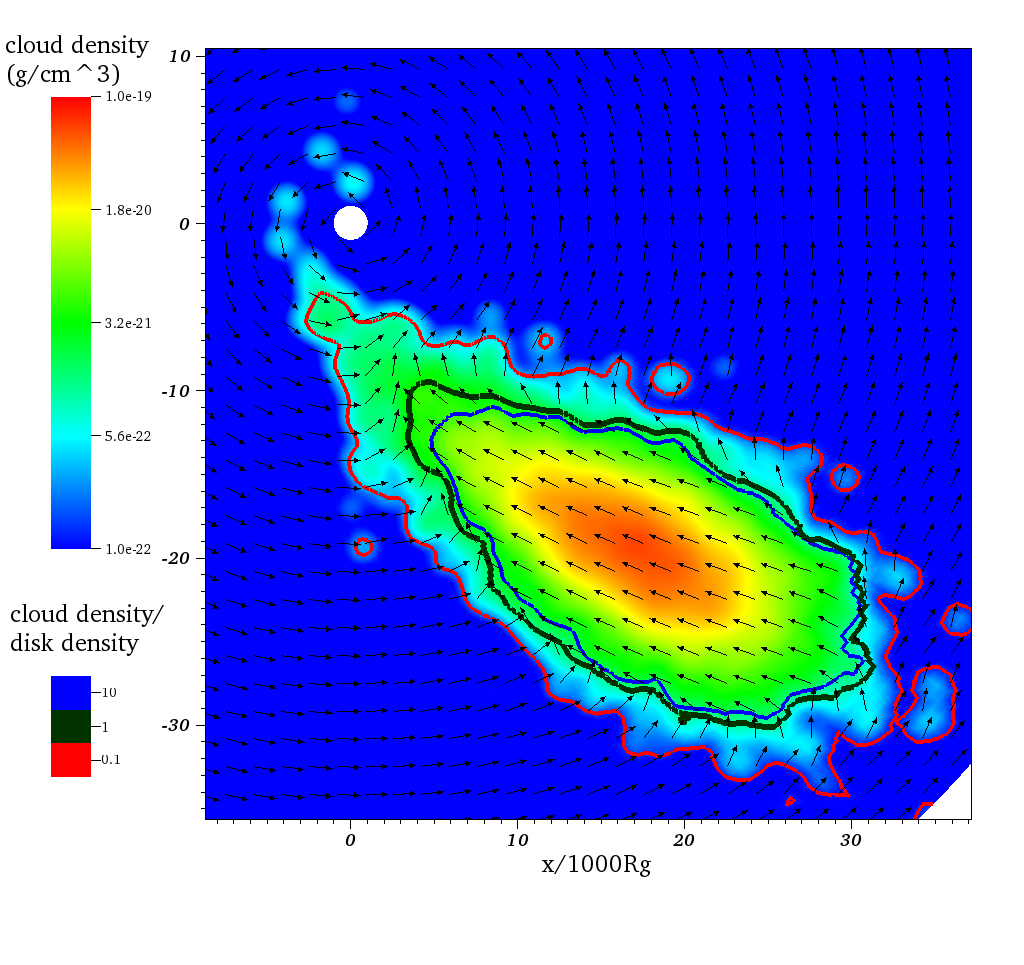}}\\\vspace{-1cm}
\subfigure{\includegraphics[width=.9\columnwidth,]{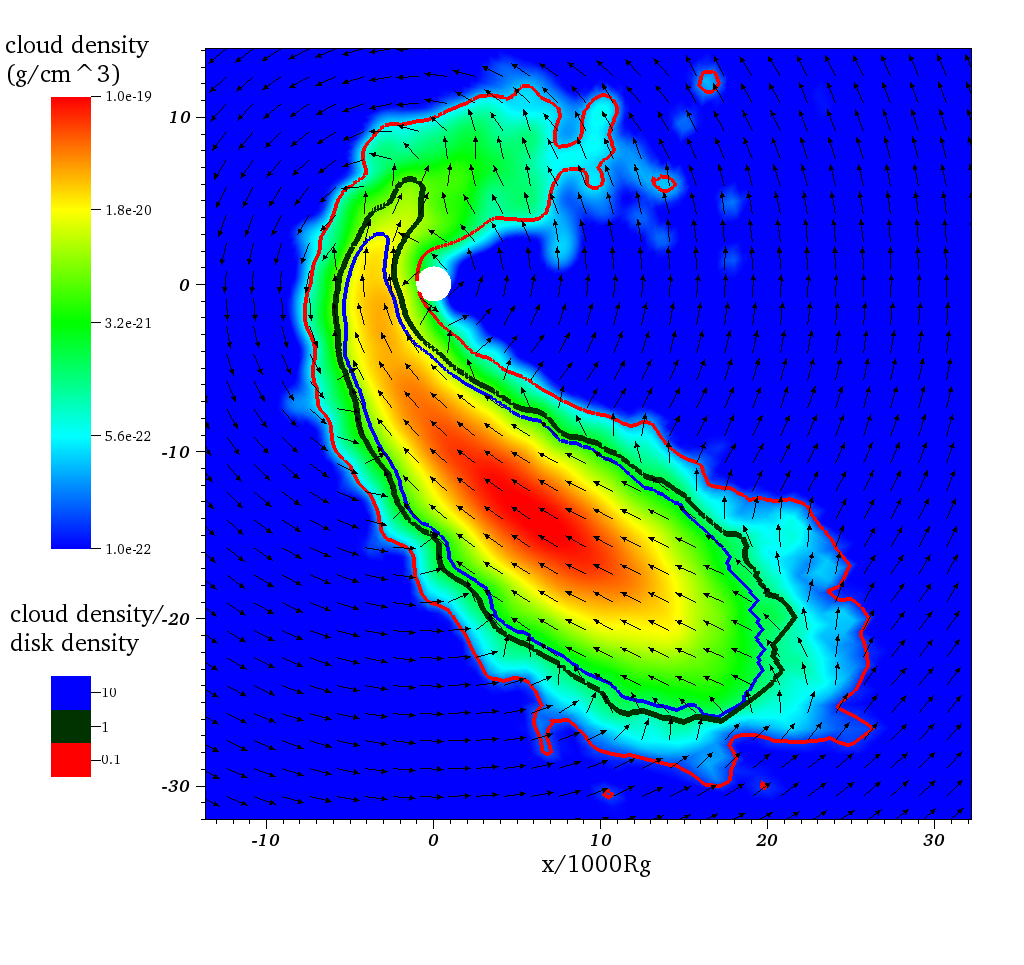}}
\caption{Density of the cloud in its orbital plane at 2012.4 (top) and
  2013.4 (bottom panel). The cloud was based on the test particle
  model as given in \citet{gillessen+13} and superimposed on top of
  counter-rotating \sgra disk . Contours show the ratio of cloud to
  disk densities. The black contour encircles the region dominated by
  the cloud density.}
  \label{f.density.gille}
\end{figure}

The test particle approach only accounts for gravitational forces, neglecting the dynamical interaction of the
cloud with the atmosphere. To test if this assumption is satisfied, we took the
cloud model as given in \cite{gillessen+13} and propagated the cloud
of test particles until 2012.4 and 2013.4, i.e., roughly until the epochs
for which position-velocity diagrams were plotted in
\cite{gillessen+12b} and \cite{gillessen+13}, respectively. We then
superimposed the implied cloud density and momentum field on top of
the background disk model. Figure~\ref{f.density.gille} shows with
colors the
cloud density distribution and contours of cloud to disk density
ratio. The black contour, in particular, surrounds the region where
the cloud density dominates. Qualitatively speaking, whatever is
outside is likely to be significantly affected by the interaction with
the disk. 

\begin{figure}
  \centering
  \begin{tabular}{MM}
  
   \begin{sideways} \large pos(mas) \end{sideways} &
\subfigure{\includegraphics[height=.2\textwidth,]{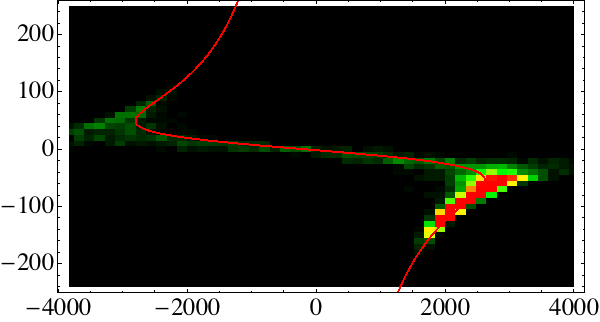}} \\
   \begin{sideways} \large pos(mas) \end{sideways} 

&\subfigure{\includegraphics[height=.2\textwidth,]{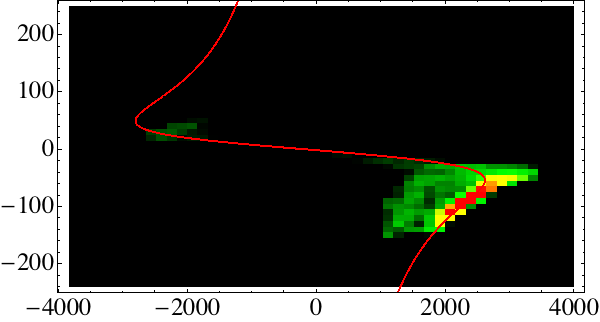}}\\
   \begin{sideways} \large pos(mas) \end{sideways} 

&\subfigure{\includegraphics[height=.2\textwidth,]{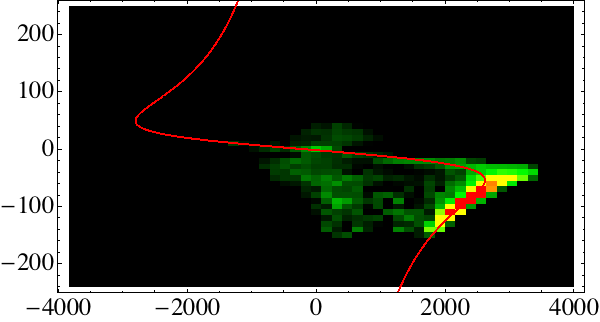}}\\
& \large LOS velocity (km/s) \\

&\subfigure{\includegraphics[width=.3\textwidth,]{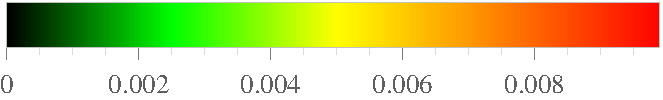}}

\end{tabular}

\caption{Position along the orbit - velocity diagrams at $2013.4$ for the cloud based
  on the \citet{gillessen+13} test particle model. The top panel
  corresponds to test particles unaffected by the disk. The middle and
  bottom panels show the position and velocities after superimposing
  the cloud model on top of counter- and co-rotating disks,
  respectively. Colors show the fractional mass of the cloud.}
  \label{f.posvel.gille}
\end{figure}

The position-velocity
diagrams shown in Figure~\ref{f.posvel.gille} quantify this
effect for $t=2013.4$. The top panel shows the line of sight velocities and positions
along the orbit for the test particle cloud model as proposed by
\cite{gillessen+13} and corresponds to the observed distributions at
this epoch. The middle and bottom panels show the cloud gas velocities
after superimposing the cloud on the counter- and co-rotating disk
atmosphere respectively. The velocities were calculated by taking a density weighted 
average between the disk and test particle cloud velocities at a given location.
The cloud gas outside the black contour is
indeed dominated by the disk and no longer moves with the original
velocity. As a result the position-velocity diagram changes. Most
striking is that there are hardly any blue-shifted (negative
line of sight velocities) particles, both for
co- and counter-rotating disks, which were directly detected in IR at
that time by \cite{gillessen+13} implying that the cloud constructed
following their model cannot reproduce the observed distributions after taking
the interaction with the disk into account.
Since we want to study the cloud properties at even later
epochs we need to construct a cloud model that after consistent
inclusion of dynamical interactions would reasonably
reproduce the features observed so far. 

Another feature of the \cite{gillessen+13} cloud model which may seem
unrealistic is the high velocity dispersion ($\sigma=120\,\rm km/s$)
assumed at $t=2000.0$. The cloud is producing line emission implying
very a low temperature ($10^4$ K) inconsistent with the adopted value of
$\sigma$. To avoid this and still reproduce the observed properties of
the cloud (size and range of line of sight velocities) one could
consider a cloud of test particles with zero velocity dispersion
intitiated at $t=2000.0$ as a spherical, constant density cloud of size
significantly larger than the corresponding characteristic length of
the \cite{gillessen+13} model \citep[similar to model F3
in][]{sadowski+G22}. The increased size would balance the lack of
velocity dispersion and produce comparable range of line of sight
velocities.
Such a flat density cloud would have much heavier outer layers which
would not be so easily affected by the disk gas. It would also be close
to isothermal what is a reasonable physical configuration for a
gaseous object like the G2 cloud.

Following this approach we have constructed a cloud model which
reasonably satisfies the constraints coming from observations up
to $t=2013.4$ (presented and discussed in Section~\ref{s.posvel}). We
adopted a flat density cloud of test particles at $t=2000.0$ with zero
velocity dispersion and radius $R_0=100$ mas. Each particle was assumed to
have the cloud center of mass velocity at that time. Test particles
were propagated until $t_{\rm init}=2012.0$ along Keplerian orbits. At that time
the cloud was smoothed using a Gaussian kernel with characteristic
width $w_{\rm kern}=1000\Rg$, scaled up to match the desired cloud
mass, and superimposed on top of the disk by summing up disk and cloud
momenta and densities. The pressure was kept at the disk-implied value
resulting in a cloud colder than the disk by 2-3 orders of magnitude, 
maintaining pressure equilibrium. \DA{We consider pressure equilibrium to
 be the most natural initial condition for 
our simulations}

\begin{figure}
  \centering
\subfigure{\includegraphics[width=.9\columnwidth,]{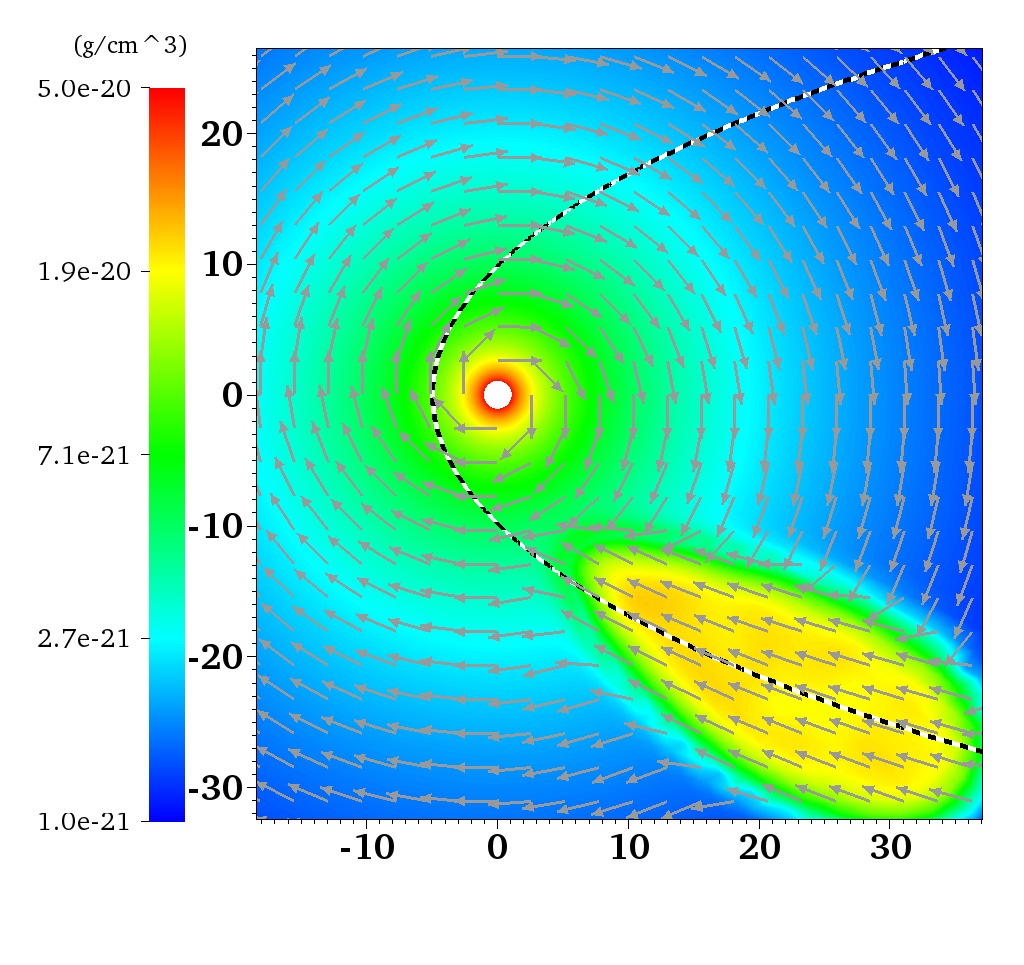}} \\\vspace{-1cm}
\subfigure{\includegraphics[width=.9\columnwidth,]{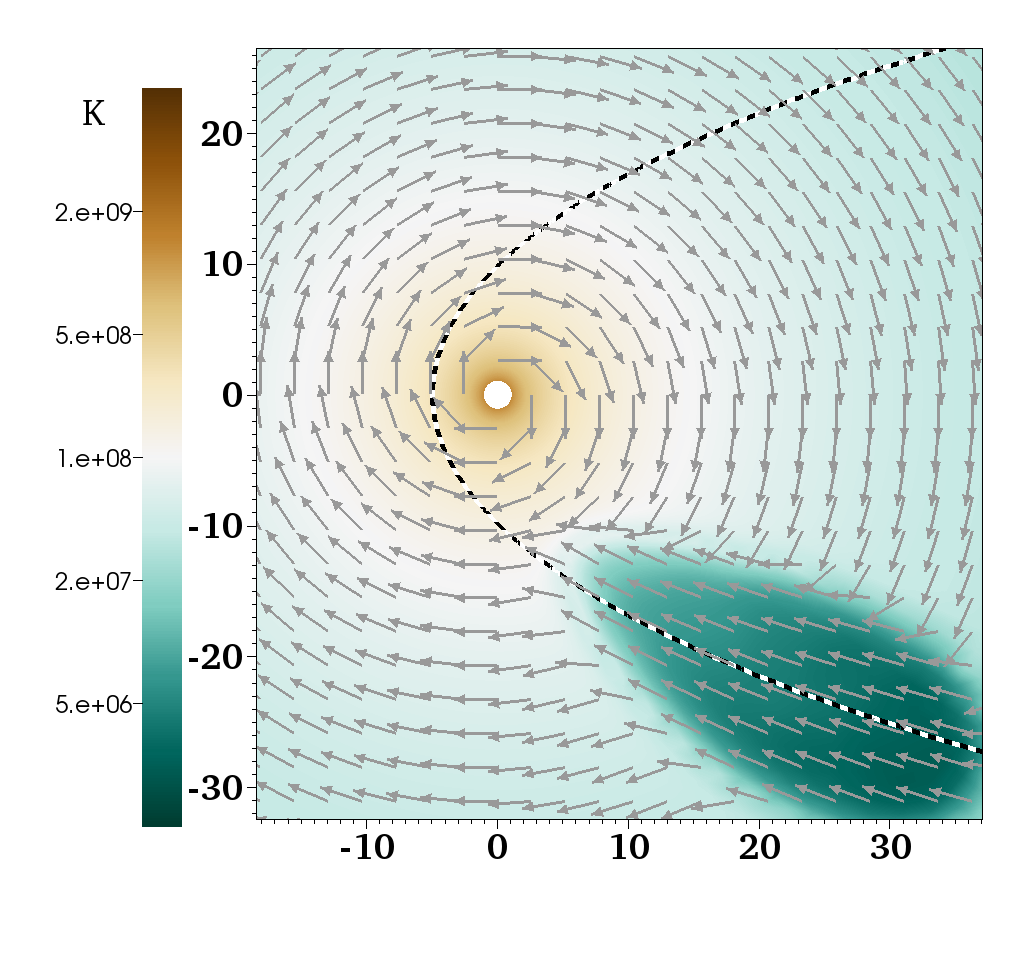}}
\caption{The initial conditions for the density (top) and temperature
  (bottom) at the cloud orbit for the fiducial model M1. Vectors show
  the velocity field and the dashed line shows the cloud center of mass orbit.}
  \label{f.initdentemp}
\end{figure}

Figure~\ref{f.initdentemp} shows the total density (top) and
temperature (bottom panel) as result from the superposition of the
cloud on to the disk atmosphere at $t_{\rm init}=2012.0$ for the fiducial model M1. Such a
setup was then used as the initial condition for the full
hydrodynamical simulation, which includes the dynamical interaction of the cloud with the accretion flow.

\section{Numerical setup}
\label{s.numerical}

We simulate the interaction of G2 cloud with the \sgra accretion disk
using the hydrodynamic part of general relativistic, radiation, hydrodynamic
code KORAL \citep{sadowski+koral}. Due to the size of the cloud orbit
and the related computational cost we were not able to resolve the BH
horizon. The inner boundary was set to $R_{\rm in}=700 R_{\rm G}$ for
most of the models\footnote{This fact makes the general relativistic
 part of KORAL redundant.}. The background atmosphere was set
according to the disk model of \cite{sadowski+G21}. On the top of it
we superimposed the cloud, as described in Section~\ref{s.cloud}, at
the time $t_{\rm init}=2012.0$ and evolved the resulting system until
and beyond the epoch of cloud pericenter $t_0=2014.25$.

The grid used in most of the simulations consisted of 160 uniformly
spaced points in the full range of azimuthal angle $\phi$, 80
quasi-logarithmically spaced points in radius between
$R_{\rm in}$
and $R_{\rm out}=50000\Rg$, and 81 cells in the polar angle $\theta$
uniformly covering the range $\theta_{0}<\theta<\pi - \theta_{0}$,
with $\theta_{0}=0.05$. We adopted
outflow boundary conditions (BC) at the inner and outer edges,
periodic BC in the azimuth and transmissive BC in the polar angle.

For all the models, the cloud orbit coincides with the equatorial plane
of the grid so that one could expect the cloud to propagate in that
plane and avoid the polar axis (where a cone with opening angle $\pm \theta_0$ has been cut off from the
domain) as much as possible. To simulate
arbitrary orientation of the cloud with respect to the disk we rotate
the background disk atmosphere by the appropriate angle.

The model of the \sgra accretion flow given by \cite{sadowski+G21} is
supposed to qualitatively reproduce the structure of the real 3D
accretion flow driven by magnetorotational instability. For obvious
reasons it does not catch all the features of the turbulent MHD
flow. It does not correspond to a hydrodynamical equilibrium torus
\citep[e.g.,][]{abramowicz+78}, either. As a result, when evolved under pure hydrodynamics,
the atmosphere is neither in hydrostatic equilibrium nor convectively stable. To
make the cloud interact with gas as similarly as possible to the
assumed disk model, we follow the approach adopted by
\cite{schartmann+12} and evolve an extra passive tracer field which
helps distinguishing the cloud-affected region from the parts of the
rotating atmosphere that have not yet interacted with the cloud. The
tracer field is initiated at $t=t_{\rm init}$ by setting its value to
the fraction of density coming from the cloud,
\be
{\rm tr}=\frac{\rho_{\rm cl}}{\rho_{\rm cl}+\rho_{\rm disk}},
\ee
and ranges from 0 (far from the cloud) to 1 (inside the cloud). Then, it is
passively evolved along the velocity field of the gas. At each time
step we consider cells with ${\rm tr}<10^{-4}$ as unaffected by the
cloud and smoothly reset the gas properties there to the original disk-model
based values.

\section{Results}
\label{s.results}

In this work, we report on eight different cloud and disk setups.
The parameters are described in Table~\ref{t.parameters}. We
place the cloud in the plane of the disk both co-rotating (M1) and counter-rotating (M2) with the disk. We incline the cloud with respect to the 
disk at 
 $60^\circ$ (M3) which produces a co-rotating model and 
 $120^\circ$ (M4) which produces a counter-rotating model. We then
take these two cases and reflect them across the cloud orbit (M5, M6).  
This does not change the dynamics of the cloud-disk interactions,
but it does change the projection of the cloud onto the plane of the sky. We then adopt two more
co-rotating clouds in the plane of disk with different cloud masses,
one at 1 $M_\oplus$ (M7, to assess the effect of a lower mass) and
one at 300 $M_\oplus$ (M8, to study ballistic-like cloud propagation).

Another parameter that defines the orientation of G2 orbit is the argument
of periapsis ($\omega$) which rotates the entire orbit around its angular momentum 
vector. When the orbit is in the plane of the disk, this has no effect on the dynamics of the system,
but when the disk is inclined, $\omega$ determines when the cloud
crosses the disk equatorial plane \citep{sadowski+G21}. 
$\omega=\pi/2$ corresponds to
pericenter lying in the equatorial 
plane of the disk. $\omega=0$ implies that the cloud crosses the disk
equatorial plane twice (at $t=t_0\pm 0.45\rm yr$) and the pericenter is
off that plane. The impact of $\omega$ on the properties of the cloud
gas after it has crossed the pericenter is secondary (the cloud plows
through gas of similar properties but at different moments of time) and we
decide to fix $\omega=0$. 

We have tested the effects of the numerical
parameters we used in our simulations. Shown in 
Table~\ref{t.parameters}, 
we ran a simulation at a higher resolution (P1), 
a simulation with a higher minimum radius (P2), a simulation
with a grid that extends further in the polar direction (P3), and two 
simulations that use higher ($10^{-3}$, P4), and lower ($10^{-5}$, P5) critical tracer values.
In all cases we observe no significant differences between our
test runs and the fiducial model M1.

We also ran one simulation to test the effect of lowering the initial temperature of our 
cloud (P6). The observed temperature has been reported to be approximately $10^4$K
\citep{gillessen+12a}, but due to the pressure equilibrium we assume, our cloud
is at a temperature of about $10^6$K. We tested the effect of lowering the initial pressure,
which was initially in equillibrium with the disk,
by one order of magnitude which caused an equal decrease in temperature, and we found 
no significant difference in the evolution of the cloud. 

\begin{table*}

\begin{tabular}{c|ccccccccccc}

\hline

Run & Disk Rotation & Inclination & Reflected & Mass ($M_\oplus$) & 
$N (R)$ & $N (\phi)$ & $N(\theta)$ & 
$R_{\rm in} (R_{\rm G}$) & $\theta_{\rm min} (\rm rad)$& $T_{\rm crit}$ & $P_{\rm cl}$\\
\hline
\hline
M1 & co & $0^\circ$ & no & 3 & 80 & 160 & 81 & 700 & 0.05 & $10^{-4}$ &$P_{\rm disk}$ \\
M2 & counter & $180^\circ$& no & 3 & " & " & " & 700 & " & " &" \\
M3 & co & $60^\circ$ & no & 3  & " & " & " & 1000 & " & " &" \\
M4 & counter & $120^\circ$ & no & 3 & " & " & " & " & " & "&" \\
M5 & co & $60^\circ$ & yes & 3 & " & " & " & " & " & " &" \\
M6 & counter & $120^\circ$ & yes & 3 & " & " & " & " & " & "&" \\
M7 & co & $0^\circ$ & no & 1  & " & " & " & " & " & " &" \\
M8 & co & $0^\circ$ & no & 300  & " & " & " & " & " & " &" \\
\hline
P1 & co & $0^\circ$ & no & 3 & 100 & 200 & 101 & 1000 & 0.05 & $10^{-4}$ &" \\
P2 & " & " & " & " & 80 & 160 & 81 & 1000 & 0.05 & $10^{-4}$ &" \\
P3 & " & " & " & "  & 80 & 160 & 81 & 1000 & 0.03 & $10^{-4}$ &" \\
P4 & " & " & " & "  & 80 & 160 & 81 & 1000 & 0.03 & $10^{-3}$ &" \\
P5 & " & " & " & "  & 80 & 160 & 81 & 1000 & 0.03 & $10^{-5}$ &" \\
P6 & co & $0^\circ$ & no & 3 & 80 & 160 & 81 & 1000 & 0.05 & $10^{-4}$ &$P_{\rm disk}/10$ \\

\hline
\end{tabular}

\caption{The parameters of the numerical models. $N(R),N(\phi),N(\theta)$ give the resolutions in $R,\phi,$ and $\theta$ respectively. 
$R_{\rm in}$ and $\theta_{\rm min}$ are the minimal radial and polar
coordinates. $T_{\rm crit}$ is the crital tracer field
value. $P_{\rm cl}$ is the initial pressure of the cloud (all but P6 are in
pressure equilibrium with the disk). }
\label{t.parameters}
\end{table*}

\subsection{Evolution of the cloud}
\label{s.evolution}

\begin{figure*}
  \centering
  \begin{tabular}{MMMM}
  \large $t = 2013.4$  & \large $t = 2014.25$ & \large$t = 2014.6$ \\  
	\subfigure{\includegraphics[height=.26\textwidth,]{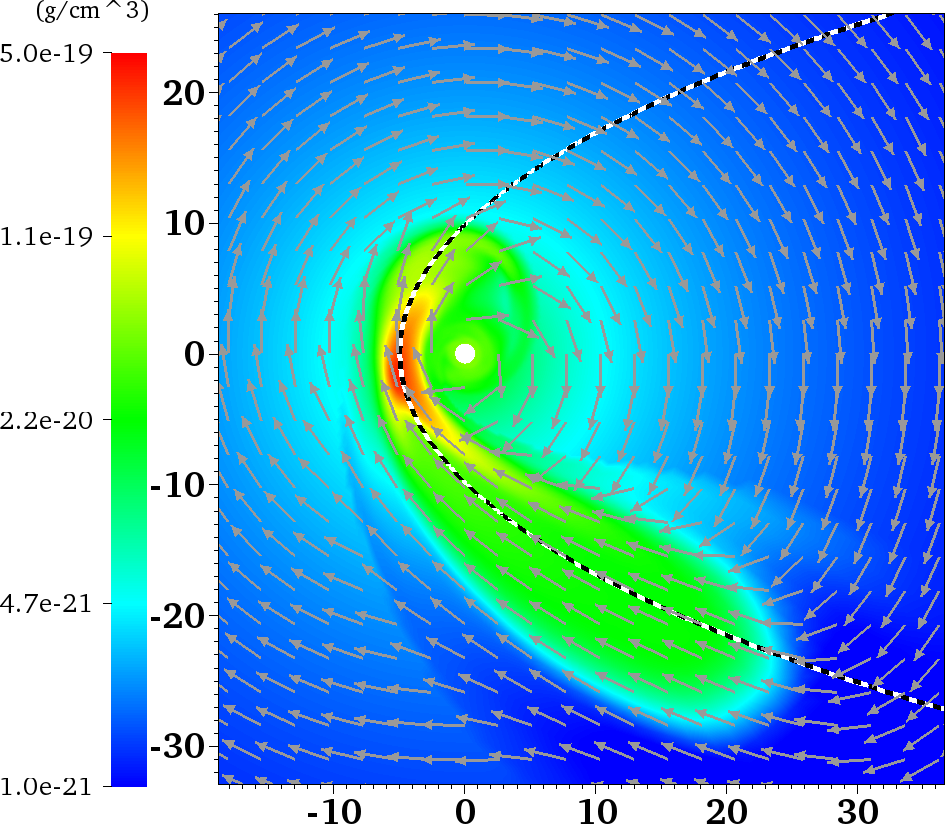}} &
	\subfigure{\includegraphics[height=.26\textwidth]{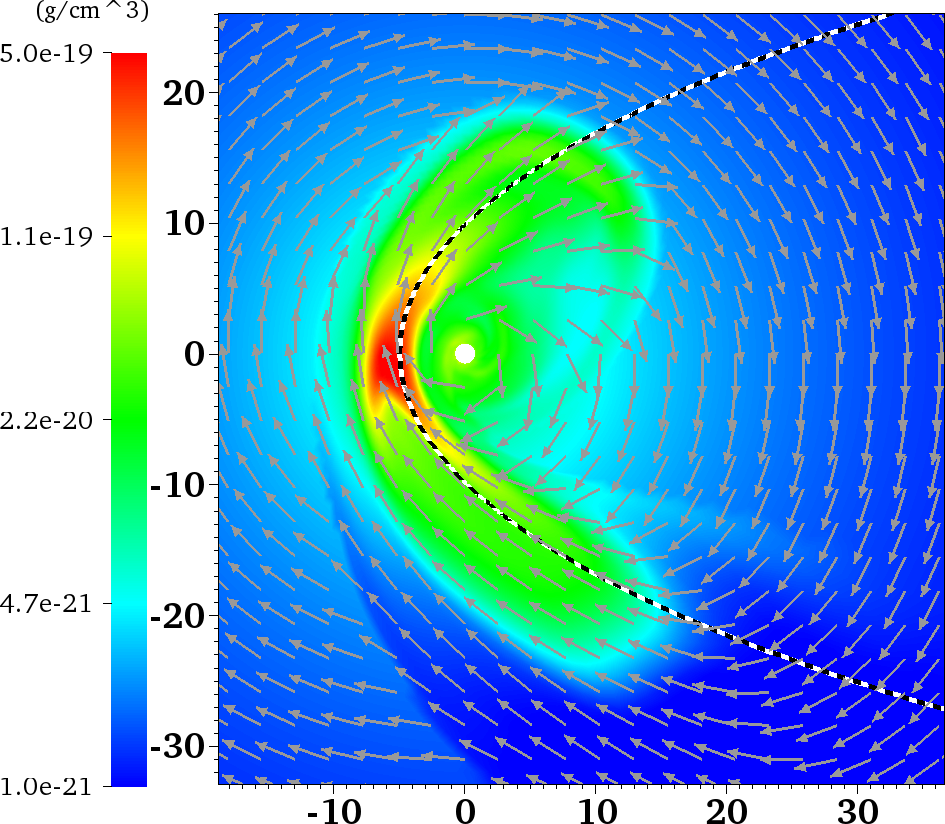}} &
	\subfigure{\includegraphics[height=.26\textwidth,]{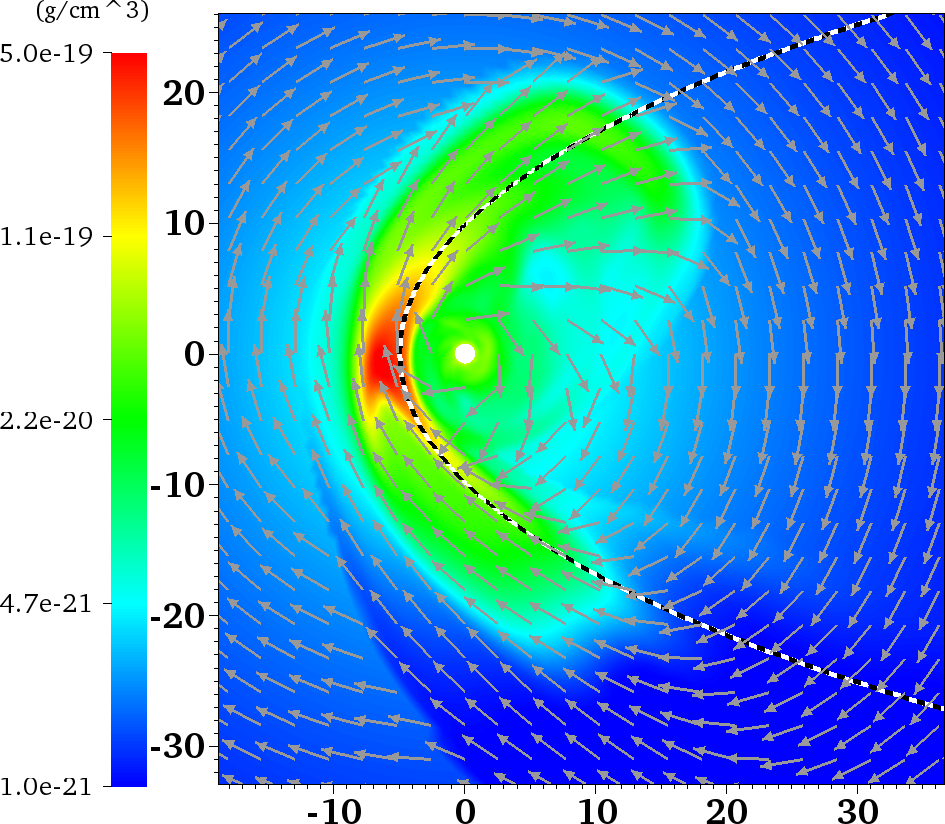}} &
	   \begin{sideways} \large $y/1000\Rg$ \end{sideways} \\
	 \large   $x/1000\Rg$ &  \large$x/1000\Rg$ & \large$x/1000\Rg$ \\
	\end{tabular}

\caption{Cloud density plots for the co-rotating disk (model M1)
  corresponding to epochs $t=2013.4, 2014.25, 2014.6$. The axes are in
  units of $1000\Rg$. The dashed line shows the center of mass orbit.}
  \label{f.density.co}
\end{figure*}

In Fig.~\ref{f.density.co} we show the evolution of the gas density in
the orbital plane of the cloud for our fiducial model (M1). Three
different epochs are presented: 2013.4 (which 
corresponds to the observations in \citet{gillessen+13}), 2014.25
(the epoch of pericenter), and 2014.6 (early August 2014).  
Overplotted on all of the images are arrows indicating the velocity unit 
vector. In this simulation, the cloud is co-rotating in the plane of the accretion 
disk, as can be seen from the velocity vectors. As the cloud approaches 
pericenter it is compressed by about two orders of magnitude before 
expanding again into the disk. The compression results mostly from
tidal forces from the SMBH. After the cloud passes pericenter, the head of the
cloud is significantly affected by the disk velocity and pulled away from the center of mass (CM) trajectory.

\begin{figure*}
  \centering
  \begin{tabular}{MMMM}
  	M1 & M2 & M8&\\
	\subfigure{\includegraphics[height=.26\textwidth]{figures/co20146density.png}} &
	\subfigure{\includegraphics[height=.26\textwidth]{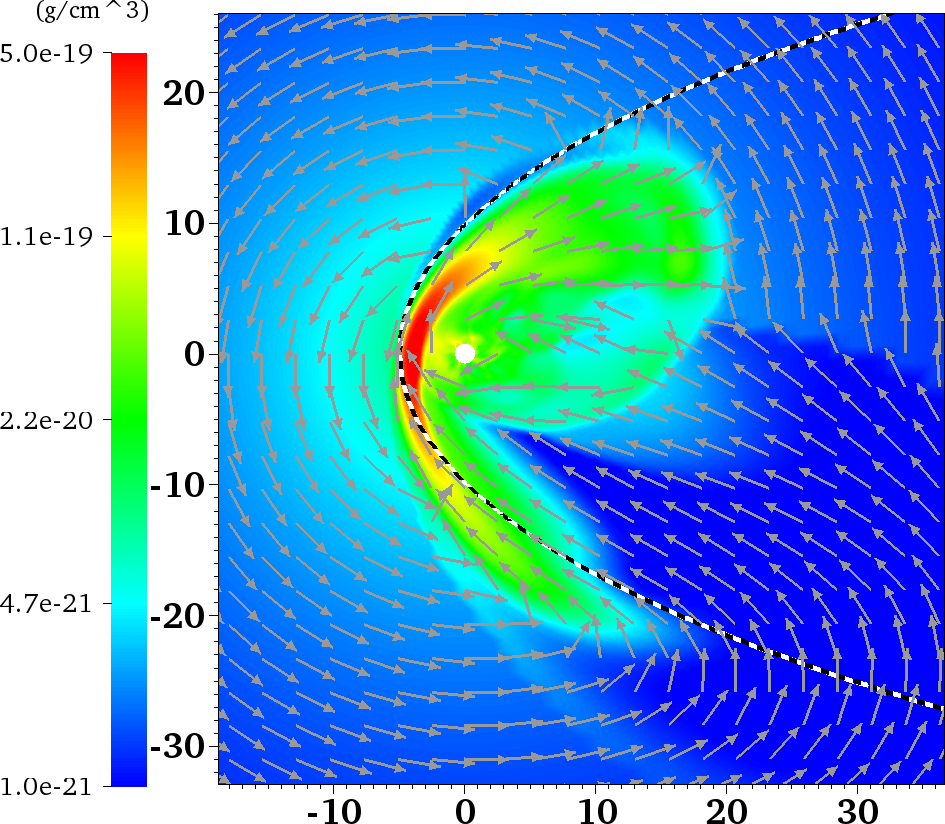}} &
        \subfigure{\includegraphics[height=.26\textwidth]{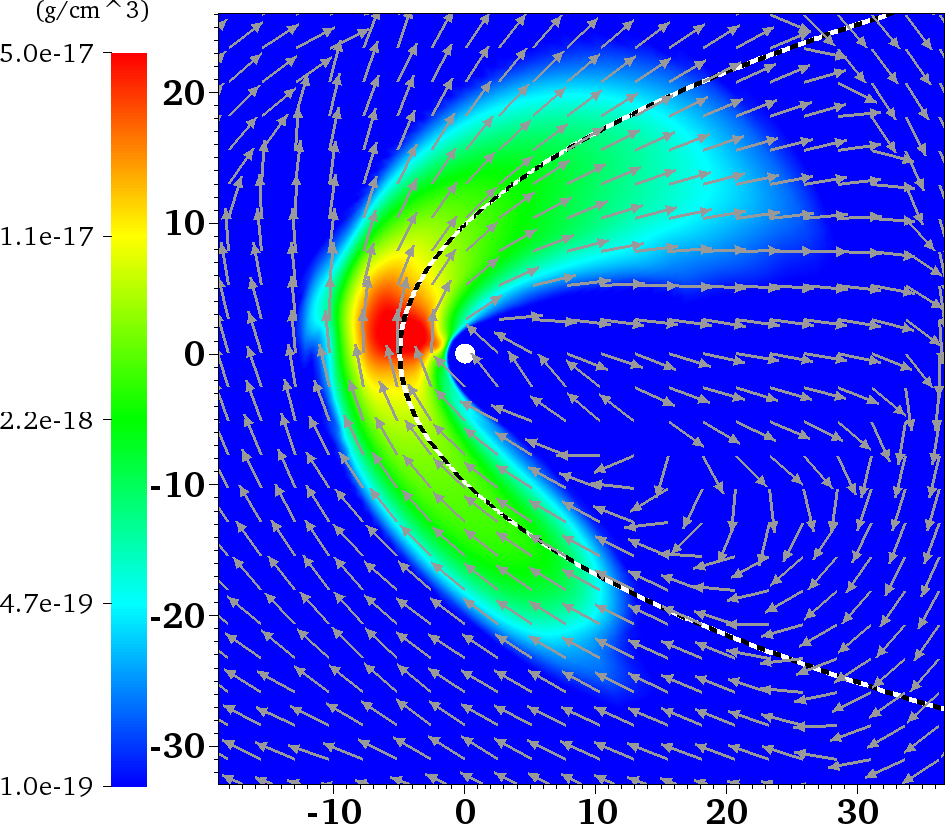}} 
        & \begin{sideways} \large $y/1000\Rg$ \end{sideways} \\
      
        M3 & M4 & M7 &\\
	\subfigure{\includegraphics[height=.26\textwidth]{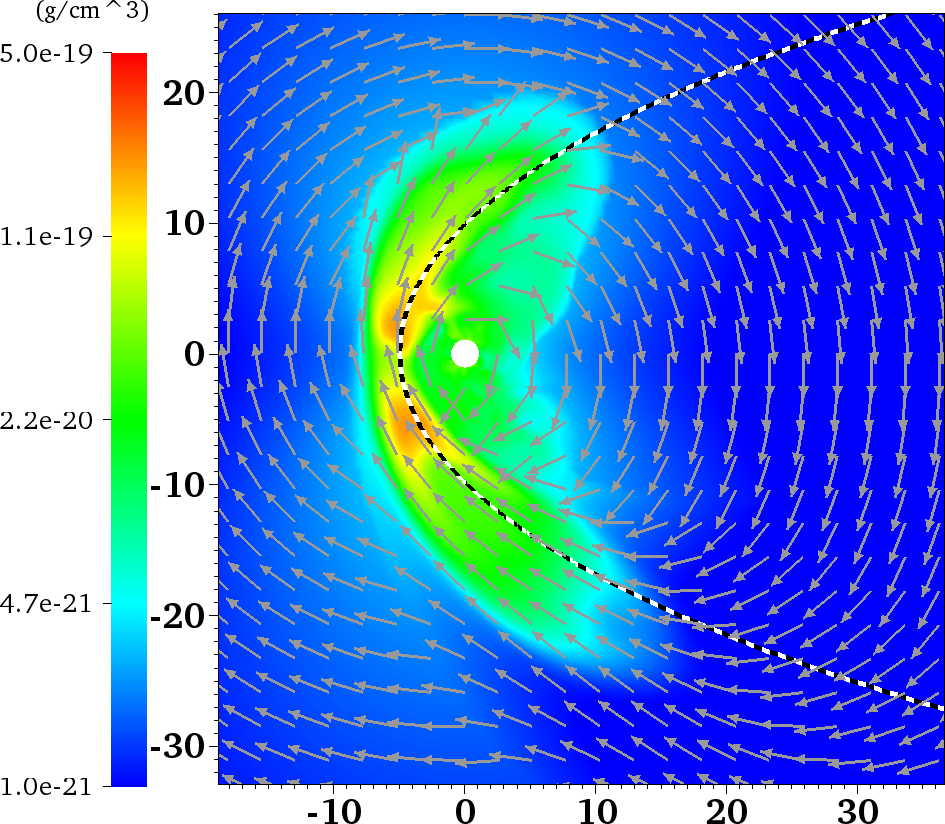}} &
	\subfigure{\includegraphics[height=.26\textwidth]{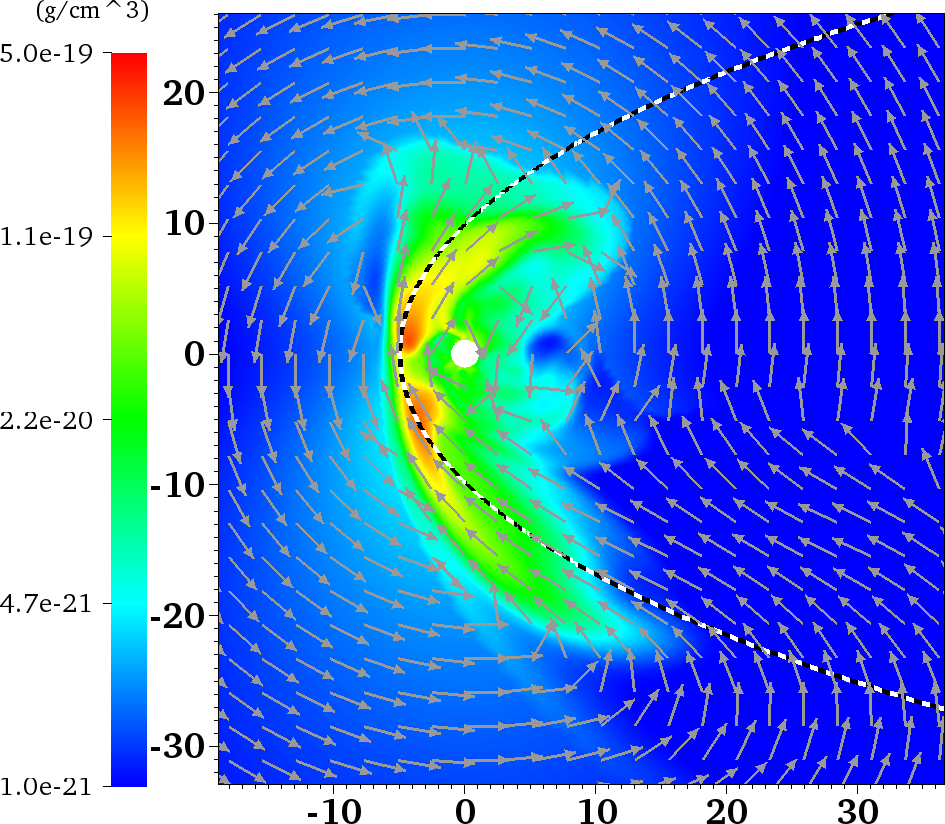}} &
	\subfigure{\includegraphics[height=.26\textwidth]{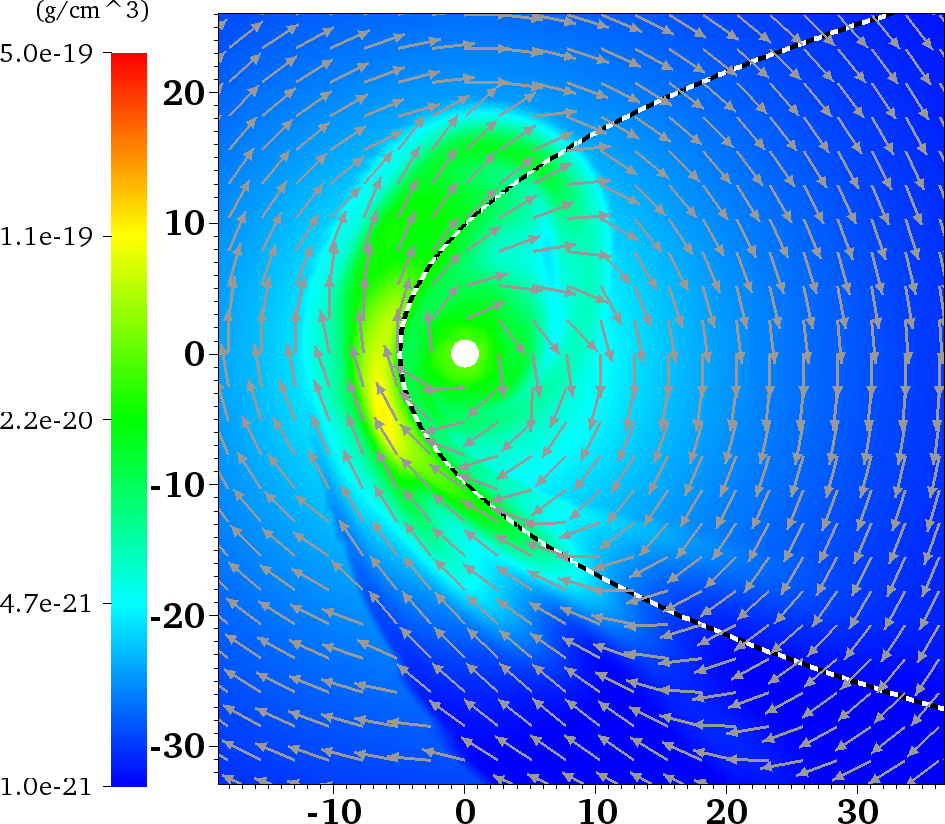}} 
	    & \begin{sideways} \large $y/1000\Rg$ \end{sideways} \\
	           \large   $x/1000\Rg$ &  \large$x/1000\Rg$ & \large$x/1000\Rg$ \\

\end{tabular}

\caption{Cloud density in the plane of orbit at $t=2014.6$. M5 and M6 are identical to M3 and M4. The axes are in units of $1000\Rg$. 
The arrows show the normalized velocity vector.}
  \label{f.density.all}
\end{figure*}

In Fig.~\ref{f.density.all}, we have shown density in the orbital plane of the 
cloud at $t=2014.6$ for six different runs. Upper left is the co-rotating run 
from Fig.~\ref{f.density.co}. Upper middle is the same except for a 
counter-rotating cloud. In this case, the bulk of the cloud passes closer 
to the SMBH, \DA{indicating a loss of angular momentum which we measure to
be approximately 30\% of the initial value by adding shells of mass at decreasing contours of density
to produce a cloud of a consistent mass between the final and initial states of the cloud and between different models. 
We then measured
the magnitude of the difference of angular momentum between the initial and final states.}
At this distance, both disk pressure and tidal forces  
are greater, and so we observe more compression. The lower left and 
lower middle plots correspond to cloud orbits inclined at
$i=120^\circ$ (co-rotating) and $i=60^\circ$ (counter-rotating), respectively.
The plots are shown in the orbital plane, and so we can see 
underdense regions at the pericenter. They result from the
disk pulling the cloud out of its orbit. It pulls it below the orbital
plane before the pericenter, and towards the opposite direction after
the pericenter. This effect is more clear in the 3D plots discussed in
the next paragraph. We see the same 
pattern of tidal and pressure compression with these two runs, except to a 
lesser extent. The right most images are both of co-rotating clouds in
plane of the disk, but for different cloud masses. Above we used a 
300 $M_\oplus$ cloud so that the cloud would be virtually unaffected by 
the disk, even near the SMBH. Below we used a 1 $M_\oplus$ cloud to 
achieve the opposite effect of emphasizing the cloud interaction with the 
disk. Note that for the 300 $M_\oplus$ cloud the density scale of the color scheme has been scaled 
with the cloud mass. 

\begin{figure*}
  \centering
    \begin{tabular}{MM}
   M1 & M2 \\
  	
 \subfigure{\includegraphics[width=.45\textwidth,]{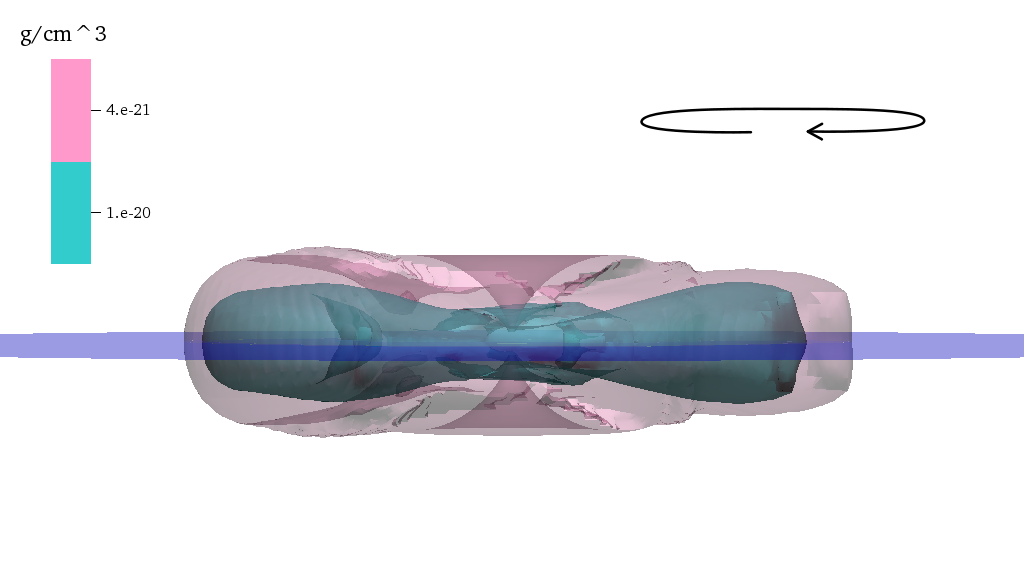}} &
\subfigure{\includegraphics[width=.45\textwidth,]{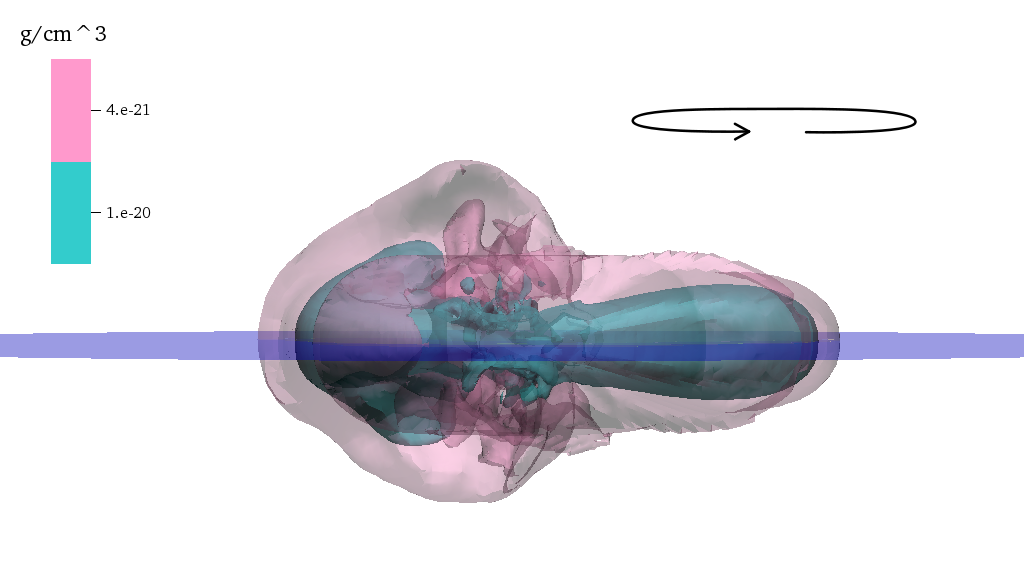}} \\
  M3 & M4 \\
	
\subfigure{\includegraphics[width=.45\textwidth,]{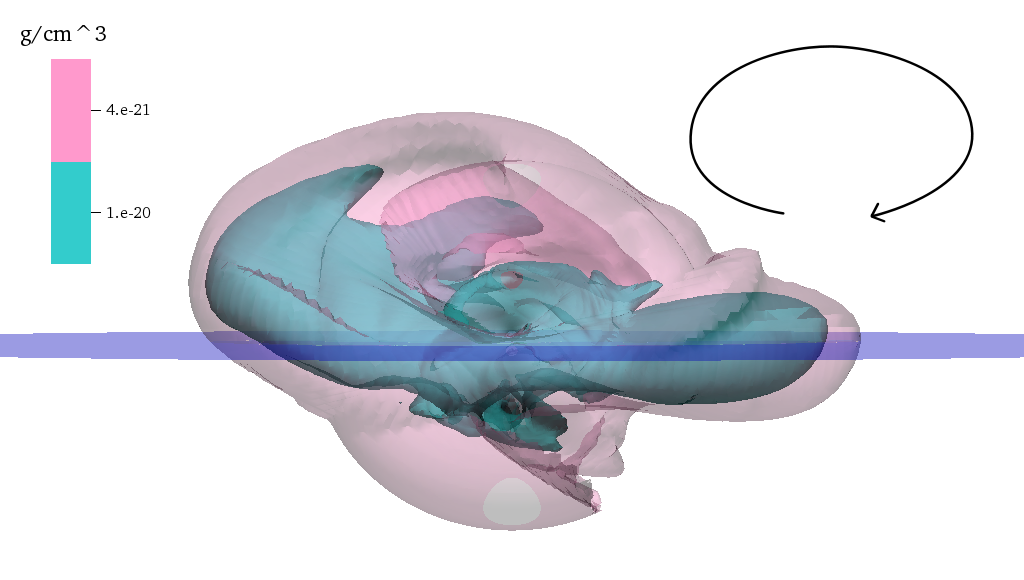}} &
\subfigure{\includegraphics[width=.45\textwidth,]{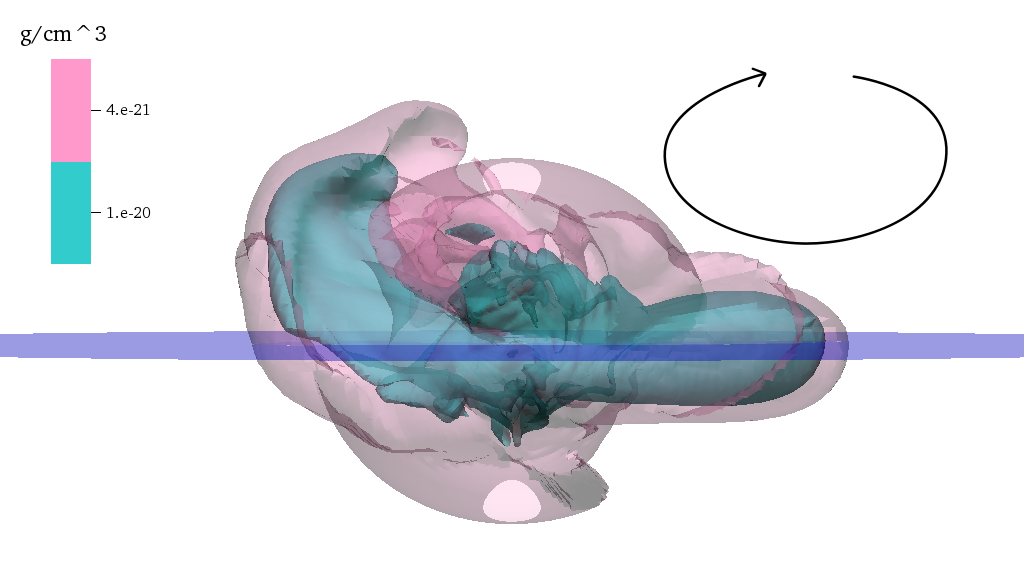}} \\
  M5 & M6 \\

\subfigure{\includegraphics[width=.45\textwidth,]{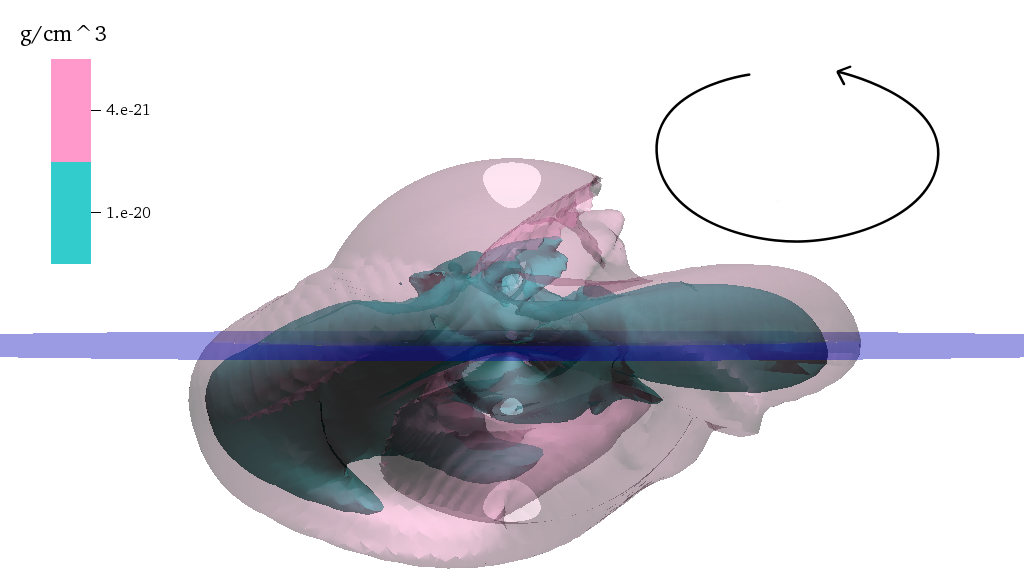}} &
\subfigure{\includegraphics[width=.45\textwidth,]{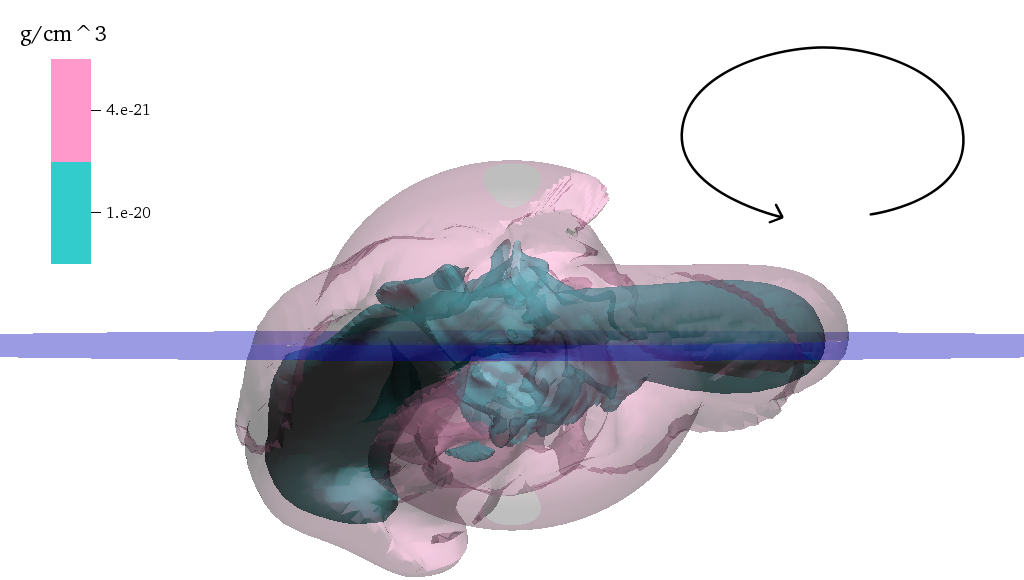}} \\

  M7 & M8 \\
\subfigure{\includegraphics[width=.45\textwidth,]{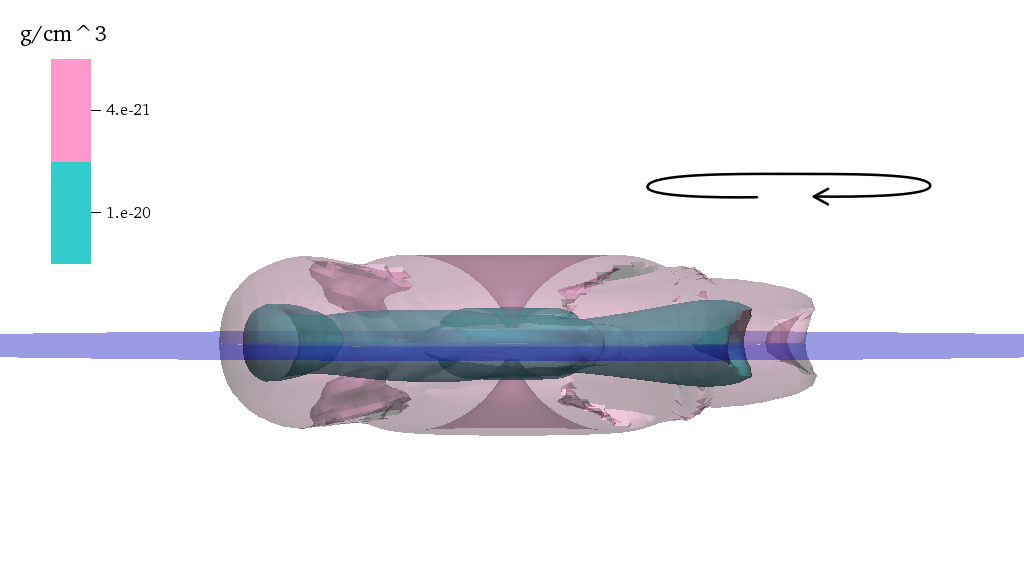}} &
\subfigure{\includegraphics[width=.45\textwidth,]{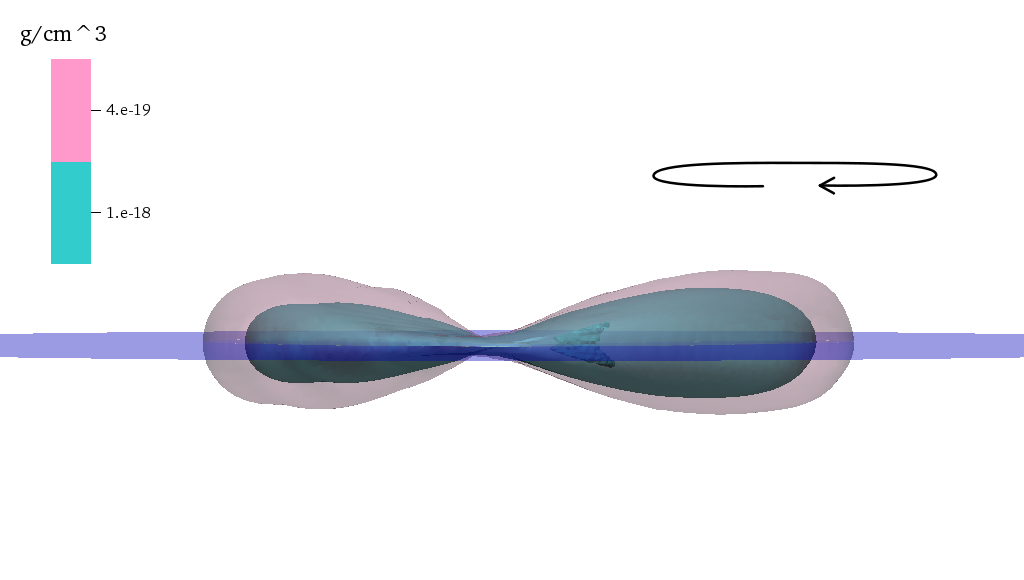}} \\

\end{tabular}

\caption{3D density contour plots at $t=2014.25$. 
The line of sight is in the cloud orbital plane from behind the location of pericenter. The blue stripe indicates the orbital plane.
 The arrows indicate the rotation and orientation of
the disk (the arrow heads are protruding). In all cases, the cloud is orbiting from right to left. The horizontal scale is 50,000 $\Rg$. The contours for model M8 were scaled by two orders of magnitude. }
  \label{f.3D}
\end{figure*}

Shown in Fig.~\ref{f.3D} are 3D
contours of density at pericenter projected along the line of
pericenter.
The orbital plane is
indicated by the blue stripe and the cloud moves horizontally to the left. 
The arrow indicates the rotation of the background accretion flow (the
arrow head is always protruding from the plot). 
The top two images are of the cloud in the plane of the
disk. We can see the cloud being disrupted but it is symmetrical with respect to the plane
of the orbit. The counter-rotating cloud shows a much stronger interaction with the disk; the head is
more strongly compressed in the direction of motion. The next two figures are for inclined orbits with
respect to the disk. In this instance, we clearly observe the disk
sweeping material out of the plane of the orbit. The next two images
are the same as the previous two, but reflected across the plane of
the orbit. We expect this effect to be observable. The last two images are of the 1 $M_\oplus$ and 300
$M_\oplus$ for which we scaled the contours in a similar way to Fig.~\ref{f.density.all}. The 1 $M_\oplus$ cloud is mostly lost
in the disk. The 300 $M_\oplus$ cloud is largely unaffected by the
disk, but highly compressed in the plane of the orbit around the
pericenter.


\subsection{Sky plane images}
\label{s.skyplane}

\begin{figure*}
  \centering
  \begin{tabular}{MMM}
    & M1 & M2 \\
    \begin{sideways} \large pos (mas) \end{sideways} &

    \subfigure{\includegraphics[width=.45\textwidth,]{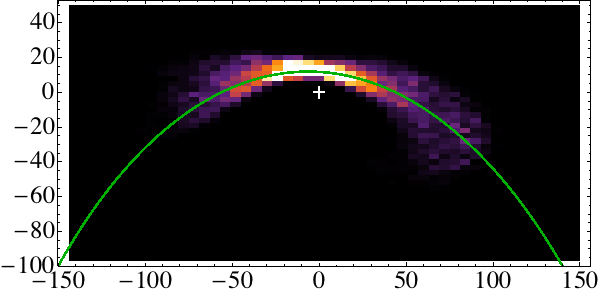}} &
    \subfigure{\includegraphics[width=.45\textwidth,]{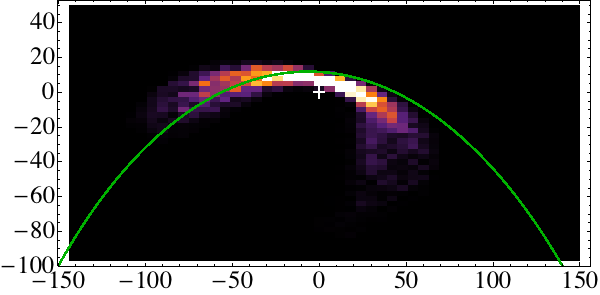}} \\
    &M3 & M4 \\
    \begin{sideways} \large pos (mas) \end{sideways} &

    \subfigure{\includegraphics[width=.45\textwidth,]{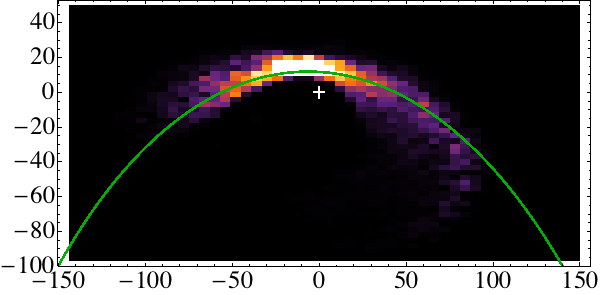}} &
    \subfigure{\includegraphics[width=.45\textwidth,]{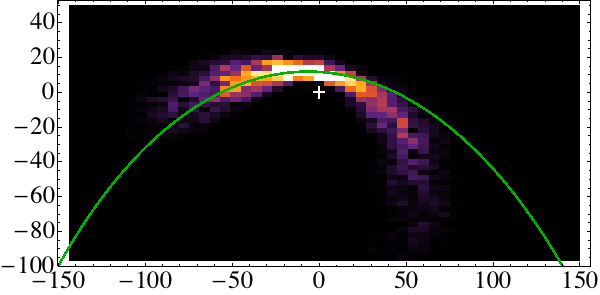}} \\
    &M5 & M6 \\
    \begin{sideways} \large pos (mas) \end{sideways} &

    \subfigure{\includegraphics[width=.45\textwidth,]{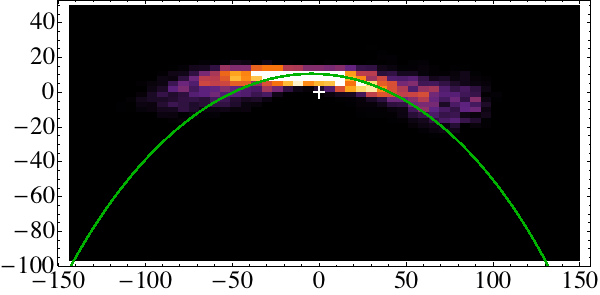}} &
    \subfigure{\includegraphics[width=.45\textwidth,]{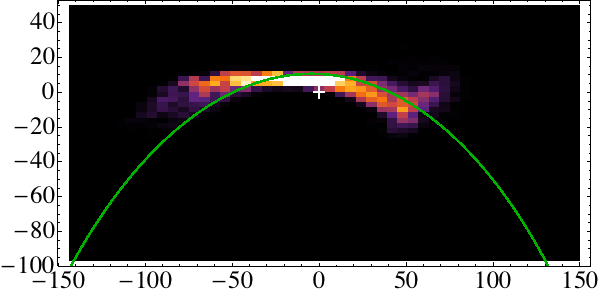}} \\

    &M7 & M8 \\
    \begin{sideways} \large pos (mas) \end{sideways} &

    \subfigure{\includegraphics[width=.45\textwidth,]{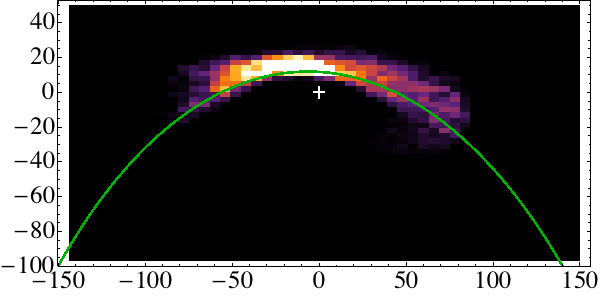}} &
    \subfigure{\includegraphics[width=.45\textwidth,]{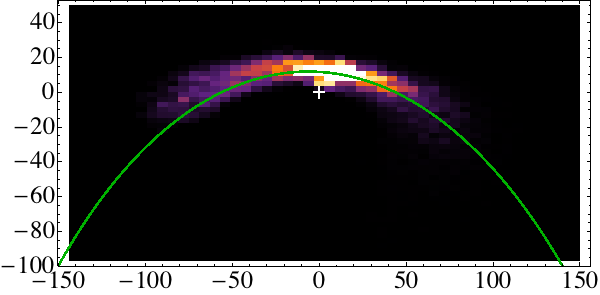}} \\
    & pos (mas) & pos (mas) \\

    &\multicolumn{2}{c}{\subfigure{\includegraphics[width=0.4\textwidth]{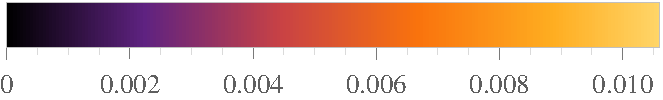}}} \\

  \end{tabular}

\caption{Plots of \DA{fractional cloud brightness on the sky} at
  the epoch $t=2014.6$.  The green line shows the trajectory of the
  center of mass projected on the plane of the sky. The axes are in
  units of mas, where Sgr $\rm A^*$ is at the origin. The colors show
  the fractional cloud mass.}
  \label{f.skydensity}
\end{figure*}

In the interest of observables, we have projected the cloud onto
the plane of the sky. We have assumed an optically thin cloud with luminosity
strictly proportional to total gas density, including the disk density which we expect to 
be dominated by the cloud. We have rotated the cloud to the
orbit orientation described in \cite{gillessen+13}. We selected all
the cells with a cutoff density of $10^{-21}\,\rm{g/cm^3}$. \DA{We binned
these cells from the simulation into a uniform grid on the sky and integrated $\rho^2$
along the line of sight,
as this is directly proportional to
flux for Case B recombination \citep{osterbrock}.  We plotted the fractional brightness on the sky,
shown in Fig.~\ref{f.skydensity}.}  We then overplotted the center of mass
orbit as projected onto the plane of the sky (green line).  

\DA{A possible caveat to consider is the effect of the temperature of the 
cloud on its observability. \cite{guillochon+14} have shown that the cooling rate
decreases with temperature, which implies a decrease in brightness after pericenter.}

We observe a clear difference
between the counter- and co-rotating cases. The former is inside its
orbit on the sky while the latter follows its orbit.  The
counter-rotating case is also denser near pericenter, which matches our
observations in Fig.~\ref{f.density.all}. For the inclined cases, we
observed in Fig.~\ref{f.3D} that material was pulled out of the plane
of the cloud orbit. We can see this effect in Fig.~\ref{f.skydensity}
where the cloud material extends about 40 mas lower on the sky. The
two inclined clouds follow the same pattern where the counter-rotating
one is farther inside the orbit than the co-rotating one. The two
images in the third row have been reflected across the plane of their orbits. This
changes their shape on the sky significantly. Both of them appear to
be in their orbits, but as we can see from Fig.~\ref{f.3D}, the bulk
of the post pericenter cloud has been pushed out of the original orbit
aligned with the observer's line of sight.  The final two images of Fig.~\ref{f.skydensity} are the 1
$M_\oplus$ and 300 $M_\oplus$ clouds.  The small cloud is 
affected and slowed down more by the disk and results in slightly more
confined cloud projection. The big cloud effectively does not feel the
disk and its front propagates faster.

\cite{gillessen+12a,gillessen+12b,gillessen+13}, using the SINFONI
spectrograph, have been able to 
spatially resolve the line of sight velocity of G2.
In this paragraph we show the plane of sky velocity distributions for
the simulations we performed without projecting the location on the
cloud orbit (as was done by these authors and what we do in the
next Section).
For this purpose we took all the cells that lie above a certain
density threshold, and binned them onto the sky in the same manner
as for Fig.~\ref{f.skydensity}. We then computed a mass weighted
average of line of sight velocity of all the cells that lie within a
particular bin and plotted them on the sky. The plots correspond to the same cloud and
disk setups as in Figs~\ref{f.3D} and \ref{f.skydensity}. Zero velocity is the same
color as the background. Deep blue color corresponds to the largest
velocities towards the observer. The velocities are the largest for
counter-rotating clouds 
which pass closer to the SMBH. The inclined and reflected runs 
show higher velocities than the original inclined models because the
velocity vectors of these clouds tend
to lie more along the line of sight.

\begin{figure*}
  \centering
 \begin{tabular}{MMM}
  & M1 & M2 \\
  	
  		\begin{sideways} \large pos (mas) \end{sideways} &
 \subfigure{\includegraphics[width=.45\textwidth,]{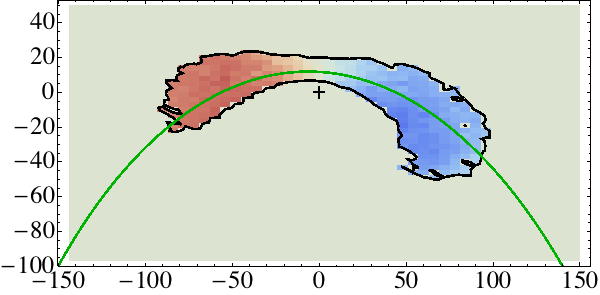}} &
\subfigure{\includegraphics[width=.45\textwidth,]{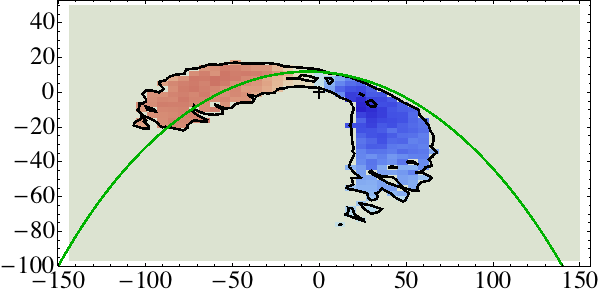}} \\
&  M3 & M4 \\
		\begin{sideways} \large pos (mas) \end{sideways} &
\subfigure{\includegraphics[width=.45\textwidth,]{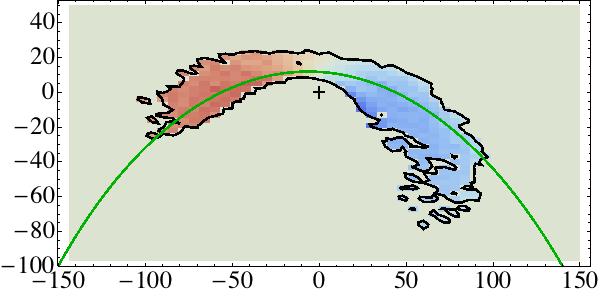}} &
\subfigure{\includegraphics[width=.45\textwidth,]{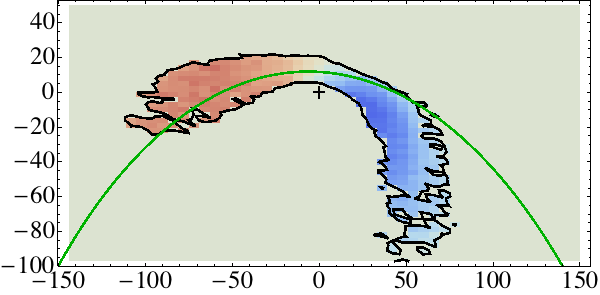}} \\
  &M5 & M6 \\
	\begin{sideways} \large pos (mas) \end{sideways} &
\subfigure{\includegraphics[width=.45\textwidth,]{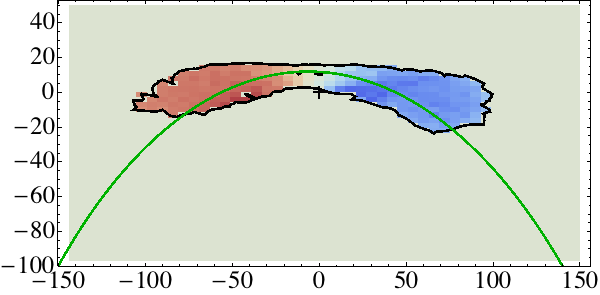}} &
\subfigure{\includegraphics[width=.45\textwidth,]{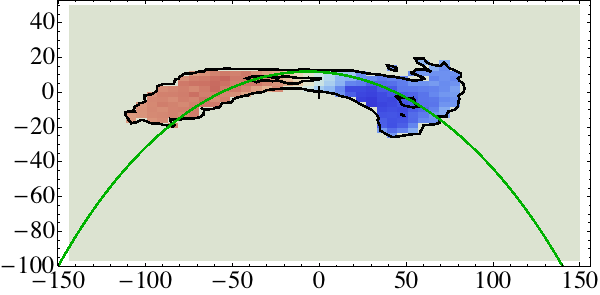}} \\

 & M7 & M8 \\
 	\begin{sideways} \large pos (mas) \end{sideways} &
\subfigure{\includegraphics[width=.45\textwidth,]{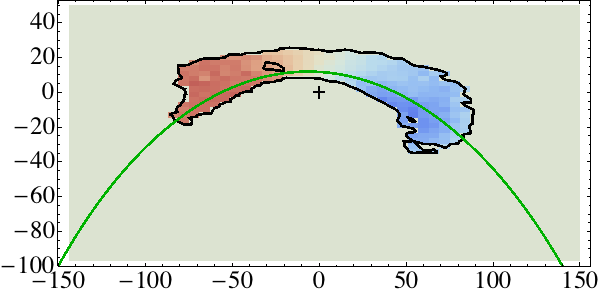}} &
\subfigure{\includegraphics[width=.45\textwidth,]{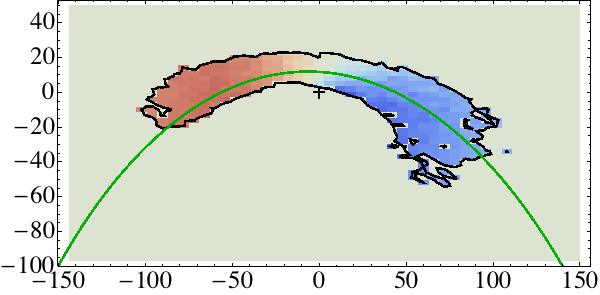}} \\
	& \large pos(mas) & \large pos(mas)\\

	 &\multicolumn{2}{c}{\subfigure{\includegraphics[width=0.4\textwidth]{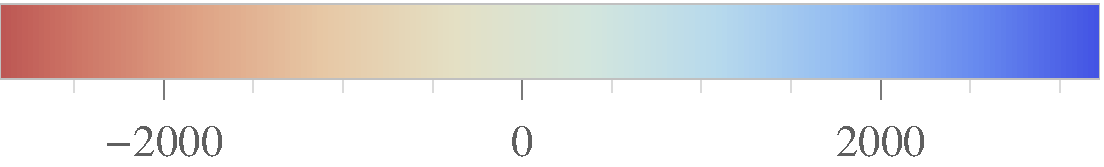}}} \\

\end{tabular}

\caption{Line of sight velocity plots at $t=2014.6$. The axes are the
  same as in Fig.~\ref{f.skydensity}. The color scale is in units of
  km/s. Only points that lie within a fractional mass contour of
  $10^{-3}$ (Fig.~\ref{f.skydensity}) were plotted. Contours show fractional mass equal $10^{-2}$ and $10^{-3}$. }
  \label{f.losvelocity}
\end{figure*}

\subsection{Position - velocity diagrams}
\label{s.posvel}

The two observables of most interest for this physical system are the position
along the orbit and the line of sight velocity. Plotting them together shows
the tidal effects of the SMBH on the cloud shape and on the velocity
dispersion \citep{gillessen+12a, gillessen+12b, gillessen+13}. We
claim that the hydrodynamical effects of the disk are also apparent
in these diagrams as deviations from the center of mass
trajectories. We use our simulation data to plot these diagrams such that they
are directly comparable with that in \cite{gillessen+13}.

 We compute the position - velocity diagrams by taking the cells above
 the same density threshold as before and binning them by their positions along the
 center of mass orbit, and their line of sight velocities. We extract
 the position along the orbit by projecting each cell location on the
 sky onto the nearest point on the projected center of mass trajectory.  
 \DA{For each bin, we integrate $\rho^2$.
 We then plot the fractional flux for each bin to indicate 
 the distribution of light from the cloud.}
 
\begin{figure}
  \centering
    \begin{tabular}{MM}
  
  	& $t = 2013.4$ \\
  	\begin{sideways} \large pos (mas) \end{sideways} &
	\subfigure{\includegraphics[width=.3\textwidth,]{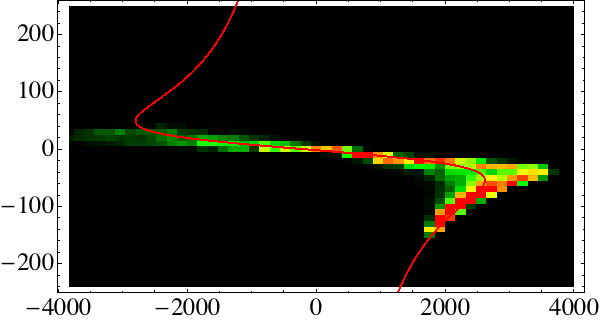}} \\
	& $t = 2014.25$ \\
		\begin{sideways} \large pos (mas) \end{sideways} &
	\subfigure{\includegraphics[width=.3\textwidth]{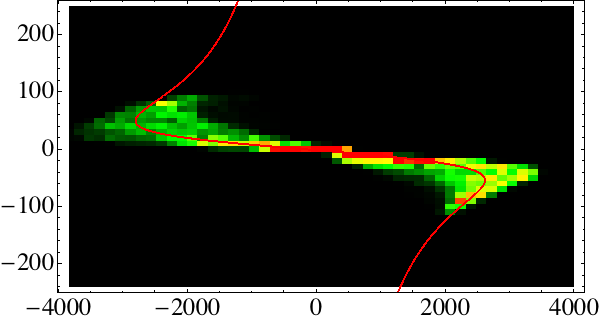}} \\
	& $t = 2014.6$ \\
		\begin{sideways} \large pos (mas) \end{sideways} &
	\subfigure{\includegraphics[width=.3\textwidth]{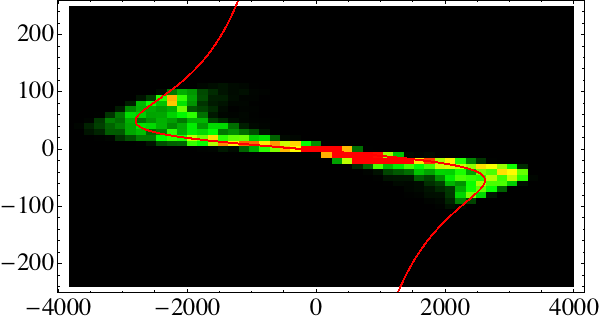}} \\
	
	& \large LOS velocity (km/s) \\
	&\subfigure{\includegraphics[width=.3\textwidth]{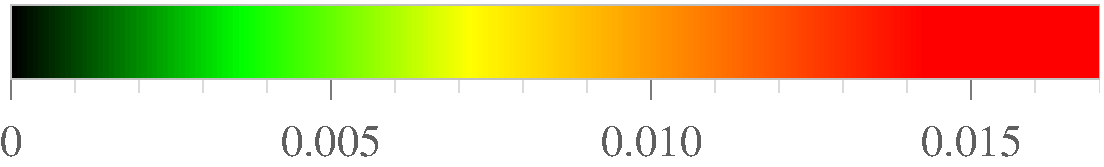}} \\
	\end{tabular}

\caption{Position-velocity plots for the co-rotating disk corresponding to epochs $t=2013.4, 2014.25, 2014.6$. The vertical axis indicates distance along the orbit from pericenter in mas. The horizontal axis is line of sight velocity in km/s. The colors indicate the fractional cloud mass.}
  \label{f.posvel.co}
\end{figure}

 In Fig.~\ref{f.posvel.co} we have produced position - velocity diagrams
 for the co-rotating cloud in the plane of the disk (model M1) at three epochs,
 2013.4 to match the observations of \citet{gillessen+13}, 
 2014.25 (pericenter), and 2014.6. At 2013.4, we have successfully
 produced some blue-shifted emission in qualitative agreement with observations. 
 We see that as the cloud
 evolves, it loses angular momentum as it interacts with the disk and
migrates towards the inside of the blue-shifted center of mass
 curve.
 
\begin{figure*}
  \centering
    \begin{tabular}{MMM}
    
    	&M1 & M2 \\
	\begin{sideways}\large pos(mas)\end{sideways} &\subfigure{\includegraphics[width=.4\textwidth]{figures/co20146posvelrhosq.png}} &
	\subfigure{\includegraphics[width=.4\textwidth]{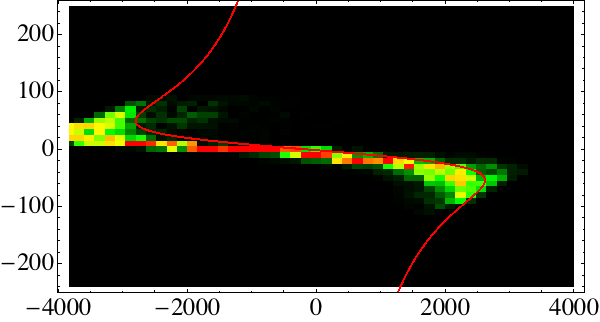}} \\
	&M3 & M4 \\
	\begin{sideways}\large pos(mas)\end{sideways}&\subfigure{\includegraphics[width=.4\textwidth]{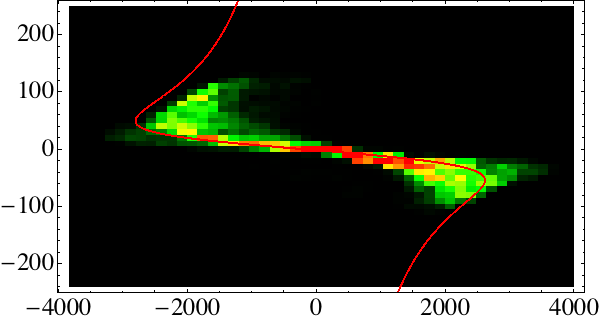}} &
	\subfigure{\includegraphics[width=.4\textwidth]{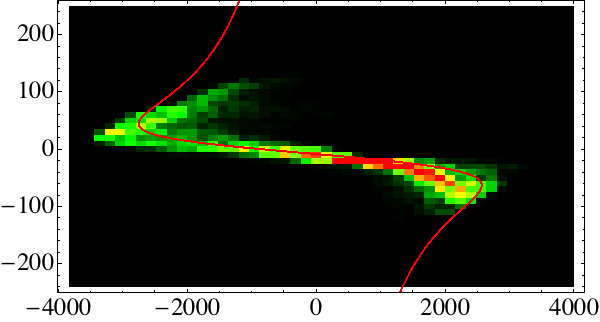}} \\
	&M5 & M6 \\
	\begin{sideways}\large pos(mas)\end{sideways}&\subfigure{\includegraphics[width=.4\textwidth]{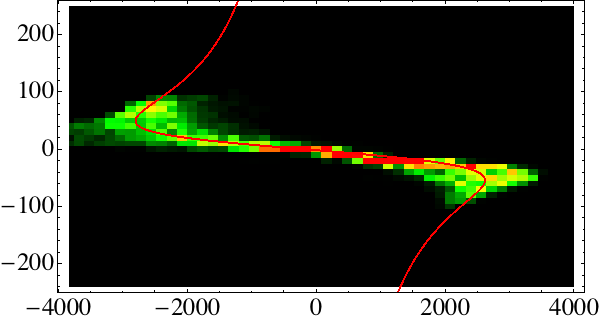}} &
	\subfigure{\includegraphics[width=.4\textwidth]{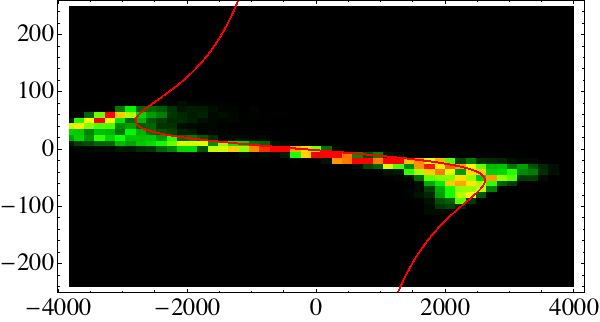}} \\
	&M7 & M8 \\
	 \begin{sideways}\large pos(mas)\end{sideways}&\subfigure{\includegraphics[width=.4\textwidth]{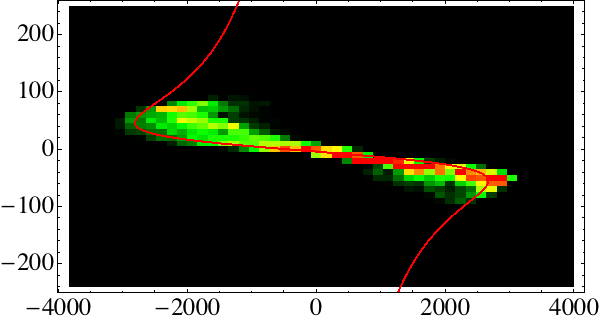}} &
	\subfigure{\includegraphics[width=.4\textwidth]{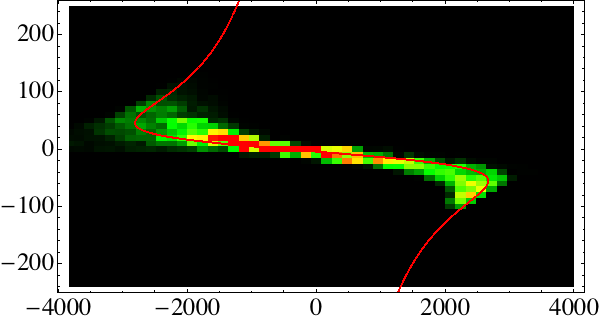}} 	\\
	& LOS velocity (km/s) & LOS velocity (km/s) \\
	&\multicolumn{2}{c}{\subfigure{\includegraphics[width = 0.4\textwidth]{figures/posvelcolorbar.png}} } 
	\end{tabular}

\caption{Position-velocity diagrams for models M1-M8 at $t=2014.6$. The vertical axis indicates distance along the orbit from pericenter in mas. The horizontal axis is line of sight velocity in km/s. The color bar scale is given in fractional cloud \DA{brightness}. The red line indicates the 
trajectory of the center of mass.}
  \label{f.posvel.all}
\end{figure*}

 In Fig.~\ref{f.posvel.all}, we have produced position velocity
 diagrams for all the runs corresponding to Fig.~\ref{f.skydensity}
 ($t=2014.6$). 
 We note that the counter-rotating run has achieved much
 faster blue-shifted velocities. From Fig.~\ref{f.density.all}, we
 argue that this is due to the cloud passing much closer to the
 SMBH. The two inclined runs seem to be less concentrated. The
 inclined, counter-rotating run shows most of its emission at lower
 blue-shifted velocities than the non-inclined run. \DA{This is due
 to the disk sweeping the gas out of the line of sight.}  In the
 reflected inclined cases, the opposite is true as seen in
 Fig.~\ref{f.skydensity}, because in such configurations the swept gas
 roughly aligns with the line of sight enhancing the radial velocity
 component. \DA{We know this is purely an effect of the line of sight,
 because there is no dynamical difference between the inclined, and the 
 inclined and reflected runs.} The reflected plots both seem to lie on the outside of the
 blue-shifted center of mass trajectory, while the unreflected plots are
 primarily on the inside.  In all six cases however, we observe the
 trend that the counter-rotating runs concentrate at higher blue-shifted velocities 
 than the co-rotating runs.  The 1 $M_\oplus$ is
 also primarily on the low velocity side of the center of mass
 trajectory, indicating it has significantly interacted with the
 disk. The 300 $M_\oplus$ cloud, however, lies precisely along the
 center of mass trajectory, indicating little or no interaction with
 the disk.

\begin{figure*}
  \centering
  \begin{tabular}{MMM}
  	
  	& \large M1 (red) M2 (blue) & \large M7 (red) M8 (blue) \\
  	\begin{sideways}\large pos(mas)\end{sideways} &
  	\subfigure{\includegraphics[height=.2\textwidth,]{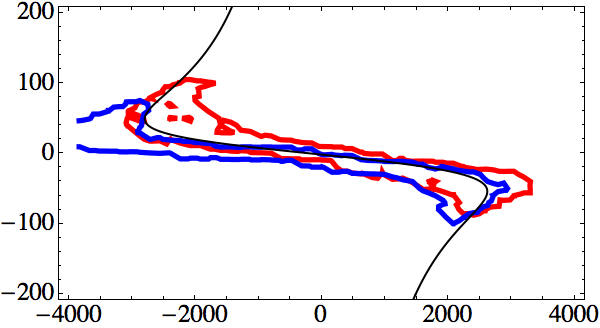}} & 
	\subfigure{\includegraphics[height=.2\textwidth,]{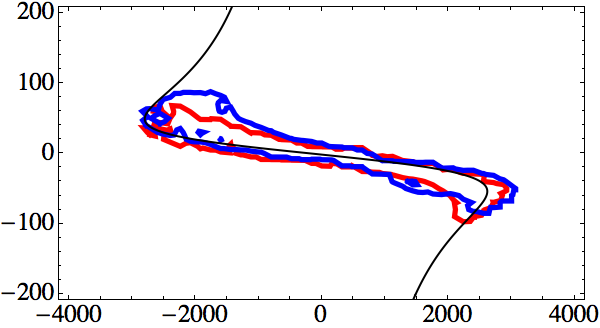}} \\

	& \large M3 (red) M4 (blue) & \large M5 (red) M6 (blue) \\
	  	\begin{sideways}\large pos(mas)\end{sideways} &

	\subfigure{\includegraphics[height=.2\textwidth,]{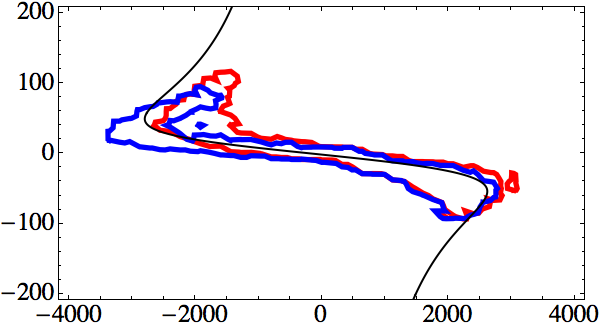}}  &
	\subfigure{\includegraphics[height=.2\textwidth,]{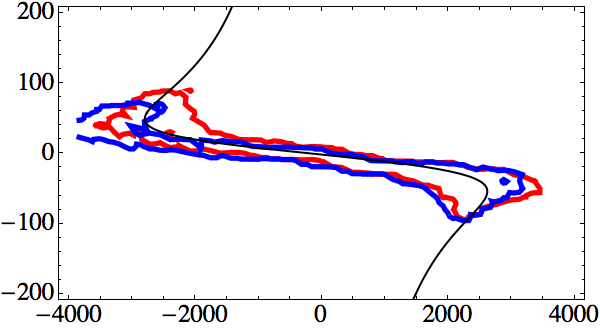}}  \\
	& \large LOS velocity (km/s) & \large LOS velocity (km/s) \\

	\end{tabular}

\caption{Contours of fractional \DA{cloud brightness (0.003) }on the position-velocity plots at $t=2014.6$. Upper left: co-rotating (red) and counter-rotating (blue) models. Upper right: 1 $M_\oplus$ cloud (red) and 300 $M_\oplus$ cloud (blue). Lower left: co-rotating and inclined (red) and counter-rotating and inclined (blue). Lower right: Same as lower left but reflected. }
  \label{f.posvelcont}
\end{figure*}

The differences between the position velocity diagrams are more
clearly shown in Fig.~\ref{f.posvelcont}, where we have plotted
contours of fractional mass on the position velocity diagrams for
different runs. Again, we see there is clear variation between the
counter- and co-rotating clouds in the plane of the disk --- the
emission from the counter-rotating cloud is more blue-shifted. We also see a clear difference
between the most and least massive clouds. The large cloud straddles the
center of mass trajectory evenly, while the small cloud contains a
significant fraction of low velocity blue-shifted emission. The bottom
two plots show how the co- and counter-rotating inclined runs appear to
lie on top of each other, and it is only when they are reflected do we
notices significant deviation.

\section{Discussion}
\label{s.discussion}

\subsection{Detectability}


Can the subtle differences between the models we considered be
detected? We have discussed the results in terms of the
position-velocity diagrams and plane of sky images. The former are
obtained using state of art adaptive optics and spectrometry. The
resolution in velocities ($\sim$75 km/s) should be enough to distinguish
between our counter- and co-rotating models. However, due to
degeneracy and uncertainties in cloud and disk parameters, only some
constraints on them may be put. Plane of sky images and spectroscopy
hold more information but the resolution required to resolve the
differences between the models ($\lesssim 0.05"$) is beyond the capabilities
of current instruments. The GRAVITY project
\citep{gillessen+gravity,kendrew+gravity} is expected to have high
enough resolution and may be useful in resolving and interpreting the
late stages of the pericenter passage.

Ultimately, what we can say about the parameters of the \sgra accretion flow
and the G2 cloud 
is limited by the resolution of future observations and, to an extent, by the
degeneracies in our models. We can, however, distinguish between groups of models.
The position velocity diagrams become less degenerate with time. Half a year after 
pericenter we expect to be able to group the models based on their similarities. 
M1, M5, and M7 all have over dense regions immediately inside the CM trajectory. M3 and M4, on average,
reach lower line of sight velocities so they are concentrated even further inside the CM trajectory.
M2 and M6 span a smaller range in positions and reach very large LOS velocities when compared to the other models.

The sky density plots add more information. They would allow us to distinguish, M1 from M5 and M7 due to the 
degree to which M1 extends below the CM trajectory in comparison. M7 seems to extend above the CM trajectory 
more so than M5, but not by much.
The difference between M3 and M4 is made slightly more clear, because the majority of M4's emission 
extends below 
the CM trajectory. M2 and M6 are very clearly different from each other; M2 has much more emission below
the CM trajectory.

The sky LOS velocity plots show slightly slower blue-shifted velocities from M7 and M5, but it is
unclear if this would actually be distinguishable.

\subsection{Assumptions}

Results presented in Section \ref{s.results} rely on a number of assumptions, as we now explain.

We assumed the G2 cloud is pressure, not momentum,
supported. Therefore, our results cannot be applied to the star wind
model of the cloud. To model the cloud we took a similar approach to
\cite{sadowski+G22} and constructed a cloud of test particles,
spherical at $t=2000.0$, but which was translated onto the numerical
grid and evolved together with the accretion flow model since
$t=2012.0.$ We chose one particular set of cloud parameters which
reasonably reproduces the observed velocities and positions at $t=2013.4$. However, this
set is not unique and other combinations of parameters may lead to
similar or better agreement with the observational constraints. We are positive
that all of them would result in cloud evolution qualitatively in
agreement with the one we studied, especially when taking into account
the uncertainty of the mass estimate \citep{gillessen+12a}.  One
should also keep in mind that the complex structure of the ``tail'',
which follows the cloud on the same orbit, suggests that the cloud
originated in a complicated process and has not at any point been
perfectly spherical.

Another set of caveats is related to the accretion flow model we
adopted. Its structure is far from constrained except in the innermost
region and close to the Bondi radius. It is tempting, and consistent
with numerical simulations, to adopt a power law interpolation between
these regions. However, nature may prefer another configuration
resulting, for example, in lower than predicted density in the intermediate
region. The accretion flow model we use is
based on a time- and spatially-averaged numerical solution and therefore
does not reproduce the expected turbulent structure. However, its
impact on the global picture of the cloud-disk interaction should be
negligible. We neglect the magnetic fields 

The numerical model we used is also limited. Most importantly, the
adopted resolution does not allow to properly resolve Kelvin-Helmholtz 
instabilities on the cloud surface \citep{schartmann+12} or 
Rayleigh-Taylor instabilities at the head of the cloud \citep{anninos+12}.
Therefore, we were able only to study the interaction of
the bulk of the cloud with the disk. The instabilities could produce
turbulent structure on the cloud surface which may enhance the
interactions. The differences between different models discussed in
Section \ref {s.results} could be, in these terms, considered
conservative. Another limitation of our numerical approach is the
necessity of stabilizing the atmosphere using the tracer
field. Such an approach may lead to under resolving the leading bow-shock
region. However, by performing simulations with different values of
the critical tracer value (Models P4 and P5) we have shown that it has
no impact on the qualitative picture.

In this work we assumed that the luminosity is proportional to the
total density, i.e., we did not distinguish the cloud density from the
density of the original accretion disk. The cloud gas is few orders of
magnitude more dense, so this assumption may have little
effect. However, we also did assume that the luminosity depends only
on the density. In reality, it may be sensitive to other parameters,
e.g., ionization fraction. To better model the observed properties of
the cloud one should perform more sophisticated radiation-transfer
simulations, similar to the ones presented in \cite{shcherbakov-g2}.

\section{Summary}
\label{s.summary}

In this work we have studied the effect of the rotating gas in the vicinity of \sgrano
on the appearance of G2 after pericenter. We have assumed
the cloud is pressure supported and adopted the self-similar model of
accretion flow by \cite{sadowski+G21}, i.e., we assumed the density
profile has a slope $r^{-1}$ between the SMBH and the Bondi radius. We
have shown that the momentum of the innermost parts of the accretion
flow is sufficient to significantly affect the cloud
evolution. Therefore, the cloud properties after it crosses the
pericenter will also depend on the gas properties of the accretion
flow.

We have performed a set of 14 hydrodynamical, three dimensional
simulations. In models M1-M6 we studied the importance of the
orientation of the cloud orbit with respect to rotation of the
disk. In models M7 and M8 we considered clouds of a lower and higher
masses. The parameter study we performed (models P1-P6) has shown our
results are to a good accuracy parameter independent.

In all the models we evolved the cloud-disk system until the epoch
t=2014.6 and studied properties of the cloud by calculating
position-velocity diagrams and plane of sky images. Various models led
to visibly different signatures. However, the differences are at the
edge of detectability for position-velocity diagrams and out of
current range for the sky-plane images. The latter may change if
Gravity \citep{gillessen+gravity,kendrew+gravity} becomes operational early enough.

The results presented in this work will help distinguish between the
pressure- and momentum-supported models of the cloud. It is also
likely they will give constraints on the nature of the \sgra accretion
flow and the G2 cloud itself.

\section{Acknowledgements}
A.S. is supported in part by NASA grant NNX11AE16G. L.S. is supported
by NASA through Einstein Postdoctoral Fellowship grant number
PF1-120090 awarded by the Chandra X-ray Center, which is operated by
the Smithsonian Astrophysical Observatory for NASA under contract
NAS8-03060. We thank Ramesh
Narayan and Maciek Wielgus for helpful discussions.

\bibliographystyle{mn2e}
\bibliography{mybib}
\end{document}